\documentclass[useAMS,usenatbib]{mnras}
\usepackage{graphicx}
\usepackage{longtable}
\usepackage{amssymb,amsmath}
\usepackage{bm}
\usepackage{multirow}
\usepackage{ulem}

\newcommand{\msol}{M$_{\odot}$}
\newcommand{\msolyr}{M$_{\odot}$\,yr$^{-1}$}
\newcommand{\hi}{H\,{\small{\sc i}}}

\newcommand{\ha}{H$\alpha$}

\newcommand{\ergs}{erg\,s$^{-1}$}
\newcommand{\ergsarc}{erg\,s$^{-1}$\,arcsec$^{-2}$}
\newcommand{\kms}{km\,s$^{-1}$}
\newcommand{\asterix}{{\rm mech}}
\newcommand{\cred}[1]{{ {\color{black} #1}}}

\newcommand{\av}{A_{V}}

\newcommand\mj{f_{j}^{k}}

\newcommand\likep{$P(F_{j}^{k} \vert \widehat{F}_{j}^{k},e_{j}^{k})$}
\newcommand\likepdf{$ (2\pi e_{j}^{k})^{-1/2} \exp{\left(\frac{F_{j}^{k}-\widehat{F}_{j}^{k}}{e_{j}^{k}}\right)^2}$}

\newcommand\priorp{$P(f_{j}^{1}f_{j}^{2}f_{j}^{3}\vert \textbf{a}$)}
\newcommand\priorpdf{$\sum_{i=1}^{N_{iso}}\,a_{i}\,\int_{M_{l,i}}^{M_{u,i}}\phi(M)\,\prod_{k=1}^{3} \mathcal{N}(f^{k}_{j}\vert \mathcal{F}^{k}_{i}(M),\sigma_i^k)dM$}

\newcommand\hpriorp{$p(\bm{a})$}
\newcommand\hpriorpdf{$\frac{\Gamma(\xi N_{iso})}{\Gamma(\xi)^{N_{iso}}}\prod_{i=1}^{N_{iso}}a_{i}^{\xi-1}$}

\usepackage[dvipsnames]{xcolor}
\usepackage{tikz,hyperref}

\definecolor{lime}{HTML}{A6CE39}
\DeclareRobustCommand{\orcidicon}{
	\begin{tikzpicture}
	\draw[lime, fill=lime] (0,0) 
	circle [radius=0.13] 
	node[white] {{\fontfamily{qag}\selectfont \tiny ID}};
	\draw[white, fill=white] (-0.0625,0.095) 
	circle [radius=0.007];
	\end{tikzpicture}
	\hspace{-2mm}
}

\foreach \x in {A, ..., Z}{\expandafter\xdef\csname orcid\x\endcsname{\noexpand\href{https://orcid.org/\csname orcidauthor\x\endcsname}
			{\noexpand\orcidicon}}
}

\title[Kiloparsec size superbubble in NGC\,628]{
The stellar population responsible for a kiloparsec size superbubble seen in the JWST ``phantom'' images of NGC\,628
}
\author[Y.\,D.\,Mayya et al.]{Y.\,D.\,Mayya$^{1\orcidB}$\thanks{Email: ydm@inaoep.mx},
J.\,A. Alzate$^{1\orcidD}$, L. Lomel\'i-N\'uñez$^{2,3\orcidJ}$, J.
Zaragoza-Cardiel$^{1,4\orcidC}$,
\newauthor  V.\,M.\,A. G\'omez-Gonz\'alez$^{5\orcidA}$, S. Silich$^{1\orcidE}$,
D. Fern\'andez-Arenas$^{2,6\orcidI}$, O. Vega$^{1\orcidG}$,
\newauthor  P.\,A. Ovando$^{1\orcidF}$,
L. H.~Rodr{\'\i}guez$^{1}$, 
D. Rosa-Gonz\'alez$^{1\orcidH}$ 
A. Luna$^{1}$, 
\newauthor
M. Zamora-Avil\'es$^{1,4}$
F. Rosales-Ortega$^{1\orcidH}$
\\
$^{1}$Instituto Nacional de Astrof{\'\i}sica, \'Optica y Electr\'onica, Luis Enrique Erro 1, Tonantzintla 72840, Puebla, Mexico\\
$^{2}$Instituto de Radioastronom\'{i}a y Astrof\'{i}sica, UNAM Campus Morelia, Apartado postal 3-72, 58090 Morelia, Michoac\'{a}n, Mexico\\
$^{3}$Universidade Federal do Rio de Janeiro, Observat\'orio do Valongo, Ladeira do Pedro Ant\^onio, 43, Sa\'ude CEP 20080-090, Brazil \\
$^{4}$Consejo Nacional de Ciencia y Tecnolog\'ia, Av. Insurgentes Sur 1582, 03940,  Mexico City, Mexico\\
$^{5}$Institute for Physics and Astronomy, Universit\"{a}t Potsdam, Karl-Liebknecht-Str. 24/25, D-14476 Potsdam, Germany\\
$^{6}$Canada-France-Hawaii Telescope, Kamuela, HI, United States\\
}

\begin{document}
\maketitle

\begin{abstract}
We here study the multi-band properties of a kiloparsec-size superbubble in the late-type spiral galaxy NGC\,628. The superbubble is the largest of many holes seen in the early release images using JWST/MIRI filters that trace the Polycyclic Aromatic Hydrocarbon (PAH)  emissions. 
The superbubble is located in the interarm region $\sim$3~kpc from the galactic center in the south-east direction. 
The shell surrounding the superbubble is  detected in \hi, CO, and \ha\  with an expansion velocity of 12~\kms, 
and contains as much as 2$\times10^7$~\msol\ of mass in gas that is mostly in molecular form.
We find a clear excess of blue, bright stars inside the bubble as compared to the surrounding disk on the HST/ACS images.
These excess blue, bright stars are part of a stellar population of $10^5$~\msol\ mass that is formed over the last 50~Myr in different star formation episodes, as determined from an analysis of color-magnitude diagrams using a Bayesian technique.
The mechanical power injected by the massive stars of these populations is sufficient to provide the energy necessary for the expansion of the shell gas. 
Slow and steady, rather than violent, injection of energy is probably the reason for the maintenance of the shell structure over the kiloparsec scale. The expanding shell is currently the site for triggered star formation as inferred from the JWST 21-$\mu$m (F2100W filter) and the \ha\ images.
\end{abstract}


\section{Introduction}

Star clusters are fundamental blocks of star formation \citep{Lada2003} and contain both low and high-mass stars whose relative numbers can be described by an Initial Mass Function \citep{Kroupa2001}.
Massive stars in clusters continuously inject energy and momentum to the surrounding interstellar medium (ISM) through stellar winds and  supernova explosions for the first 40~Myr \citep[][STARBURST99 henceforth]{SB99}.
The injected energy can destroy the natal cloud (negative feedback), and/or trigger star formation in a neighbouring cloud (positive feedback), or under some circumstances modify the structure of the surrounding ISM into rings which are referred to as holes, cavities, bubbles, superbubbles, shells etc. \citep{Tenorio-Tagle1988}. We will refer to these structures as bubbles in the present work. The role of stellar feedback in the creation of bubbles has been numerically and analytically studied by several groups \citep[e.g.][]{MacLow1988,DeYoung1994,MacLow1999,Silich2001,1992ApJ...388...93K, 2020MNRAS.495.1035S}.

Bubbles are most easily traced as HI holes by the interferrometric HI observations \citep{Brinks1986,Boomsma2008}. They are also traced in the ionized gas as shells or filaments \citep[e.g.][]{Hunter1990}. Late-type spirals and irregular galaxies are known to show a rich population of bubbles, with their radius varying from tens of parsecs to a kiloparsec \citep[see e.g.][]{Pokhrel2020}. The number of bubbles in galaxies scales with the star formation rate (SFR) supporting the basic idea that most of the bubbles are created by the feedback from massive stars. Holes occupy as much as 15\% of the surface area traced by \hi\ in some late-type galaxies \citep[][]{Pokhrel2020}, giving them a porous appearance. 

Stellar feedback is found to provide enough energy to explain the observed sizes of bubbles \citep[][]{Vorobyov2005}. However, systematic searches for the remnant cluster that created bubbles have  failed to detect an obvious candidate \citep[][]{Warren2011}. These authors conclude that the \hi\ holes are likely formed from multiple generations of star formation and only under suitable interstellar medium conditions. 

The MIRI camera mounted on the JWST provides a new tool to study the formation of bubbles \citep[][]{Bouchet2015,Wright2015,Wells2015}.
The F770W filter intercepts the Polycyclic Aromatic Hydrocarbon (PAH) features, which traces the Photo-Dissociated Region (PDR) phase and the diffuse medium associated with the neutral gas \citep[e.g.][]{Tielens2008, Montillaud2013, Hensley2022}.
\cred{
The PAH molecules are most commonly excited by soft ultraviolet radiation from stars \citep{Tielens2008}.} The PAH molecules can be easily destroyed by hard radiation either from massive stars or an Active Galactic Nucleus (AGN) and high-velocity shocks from jets and or SNe explosions \citep{Micelotta2010a, Zhang2022}. 
\cred{Collision with electrons in pre-shock hot gas also destroys the PAH molecules \citep{Micelotta2010b}. However, PAH molecules are known to survive in the shells surrounding the expanding bubbles \citep[see e.g.][]{Pabst2019, Pabst2021}.}

We here use the images of NGC\,628 from Early Science data release of JWST principally in the F770W filter to detect a large bubble of $\sim$1~kpc diameter. The bubble is also traced in the F1000W, F1130W and F2100W filters that trace Silicate features, large PAHs and the hot dust continuum, respectively. The bubble studied in this work is the largest of numerous bubble-like structures in the F770W of this galaxy, which has given rise to the nickname ``phantom galaxy'' for galaxies that have porous appearance in this filter. NGC\,628 is the first of such galaxies for which data are released. 
We complement the JWST data with the publicly available optical images 
from the HST/ACS filters and nebular line images from the MUSE spectral datacube. The detailed study of the largest bubble that we carryout here would help to address the correspondence between cluster feedback and bubble formation. NGC\,628 is a late type spiral galaxy of solar metallicity \citep[][]{Kreckel2019} at a distance of 9.77~Mpc \citep[][\cred{47~pc\,arcsec$^{-1}$}]{Olivares2010}. It is mildly star-forming with a global star formation rate of 1.7~\msolyr\  \citep[][]{Santoro2022}.

Section 2 describes the data used in this study. Sec.~3 presents the morphology, kinematics and ionization state of the ionized gas associated with the bubble using the \ha\ line. In Sec.~4, we present the results about the stellar population responsible for creating the bubble. We address the energetics, star formation scenario and the special circumstances that create kiloparsec bubbles in Sec.~5.

\begin{figure}
\begin{centering}
\includegraphics[width=0.99\linewidth]{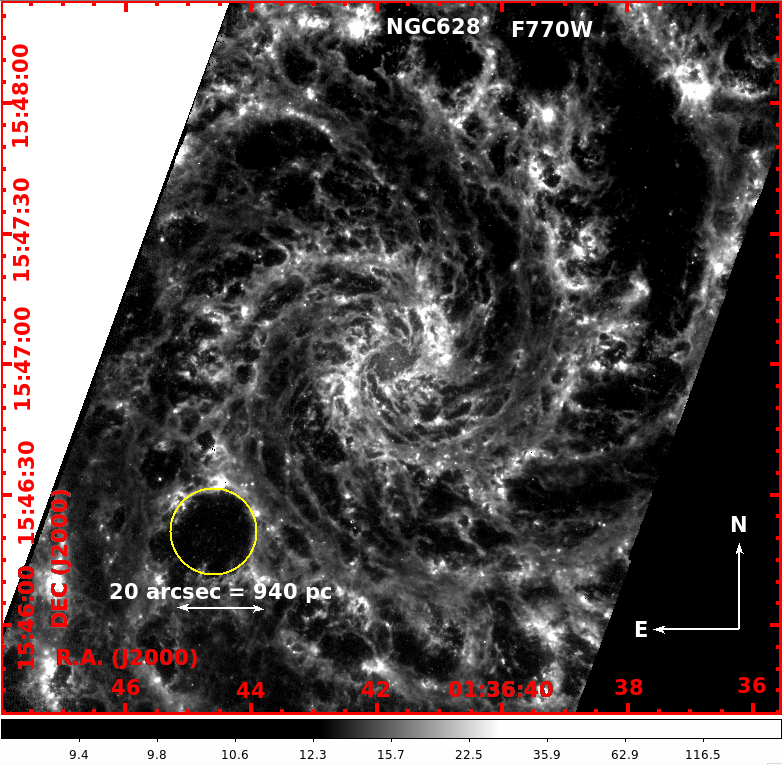}
\par\end{centering}
\caption{A giant bubble in NGC\,628 traced by the PAH emission in the filter F770W of JWST. The location of this cavity, just outside the spiral arm to the south-east of the nucleus, is indicated by a yellow circle of 20 arcsec in diameter, which corresponds to 940~pc at the distance of 9.77~Mpc to NGC\,628. \cred{A gray-scale bar is given at the bottom, where the numbers are in units of MJy/sr.}}
\label{fig:bubble}
\end{figure}

\section{JWST observations of NGC\,628 and the detection of the bubble}

\begin{table*}
\small\addtolength{\tabcolsep}{-3pt}
\caption{Description of imaging data used in this work}
\label{tab:data}
\begin{tabular}{llllllllll}
\hline
Telescope/ &  Filter  & $\lambda_{\rm c}$ & $\Delta\lambda$ & \cred{Pixel} & \multicolumn{2}{c}{FWHM}         & ZP & m$_{\rm lim}$ & tracer \\
instrument       &          &   $\mu$m          & $\mu$m      &\arcsec\    & \arcsec & pc          &   mag & mag &   \\
    (1) &  (2)     &            (3)    &    (4)          &  (5)   &   (6)           &  (7) & (8) & (9) \\
\hline
HST/{\sc acs} &  F435W  & 0.43  & 0.10 & 0.05 & 0.10 & 4.70 & 25.79 & 28.06 & stars \\ 
HST/{\sc acs} &  F555W  & 0.53  & 0.12 & 0.05 & 0.10 & 4.70 & 25.73 & 27.33 & stars \\ 
HST/{\sc acs} &  F814W  & 0.83  & 0.25 & 0.05 & 0.10 & 4.70 & 25.53 & 26.47 & stars \\ 
JWST/{\sc nircam} &  F200W  &  1.99  & 0.22 & 0.031 & 0.06 & 3.01 & 26.38 & 25.21 & stars \\ 
JWST/{\sc nircam}   &  F300M  &  2.99  & 0.32 & 0.063 & 0.13 & 6.11 & 24.05 & 23.47 & stars \\ 
JWST/{\sc nircam}   &  F335M  &  3.36 & 0.35 & 0.063 & 0.13 & 6.11 & 23.82 & 23.37 & stars\\ 
JWST/{\sc nircam}   &  F360M  &  3.62  & 0.39 & 0.063 & 0.13 & 6.11 & 23.67 & 23.32 & stars \\ 
JWST/{\sc miri} &  F770W  &  7.7  & 2.20 & 0.11 & 0.25 & 11.8 & --- & --- & PAH \\ 
JWST/{\sc miri} &  F1000W & 10.0  & 2.00 & 0.11 & 0.32 & 15.0 & --- & ---  & Silicate\\ 
JWST/{\sc miri}    &  F1130W & 11.3  & 0.70 & 0.11 & 0.36 & 16.9 & --- & --- & PAH \\ 
JWST/{\sc miri}    &  F2100W & 21.0  & 5.00 & 0.11 & 0.67 & 31.5 & --- & --- & hot dust\\ 
\hline
\end{tabular}
\end{table*}

\begin{table}
\small\addtolength{\tabcolsep}{-3pt}
\caption{Datacubes used for kinematical and line maps analysis}
\label{tab:datacube}
\begin{tabular}{lllllllll}
\hline
Telescope &  line & $\lambda$ & $\Delta$V &  \cred{Pixel} & \multicolumn{2}{c}{BEAM} & tracer \\
/instrument         &         &         &  km/s  & \arcsec  & \arcsec & pc          & \\
    (1)  &  (2)    &   (3)   &    (4)    &  (5)   &   (6)           & (7) & 8 \\
\hline
VLT/{\sc muse}   & \ha\  & 6563~\AA\  & $\sim$15    &   0.2 &         0.6    & 28 & H$^+$\\ 
ALMA/{\sc phangs} &  CO & 1.3~mm   & 2.5 & 0.2 & 2 &  94 &  H$_2$\\ 
VLA/{\sc things} &  H{\sc i}  & 21~cm  & 2.6& 1.5 & \cred{ 6.9}$\times\cred{ 5.6}$ & \cred{300} & \hi\ \\ 
\hline
\end{tabular}
\end{table}

\begin{figure*}
\begin{centering}
\includegraphics[width=0.48\linewidth]{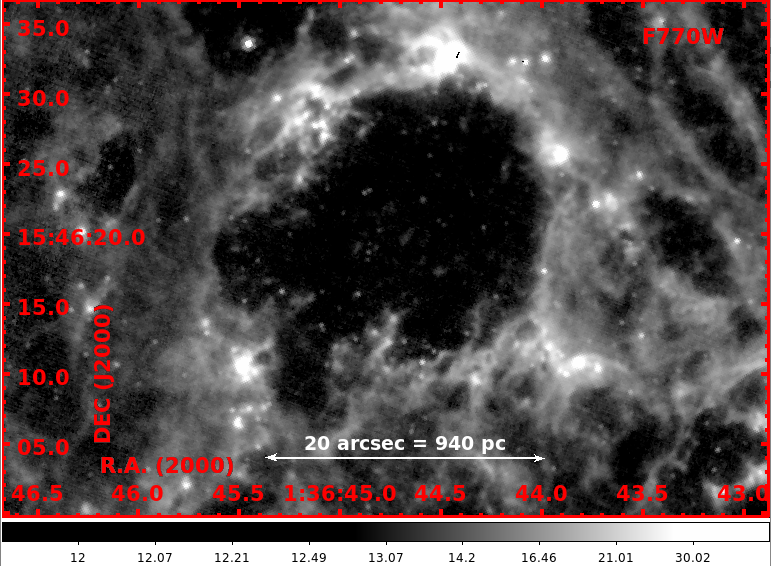}
\includegraphics[width=0.48\linewidth]{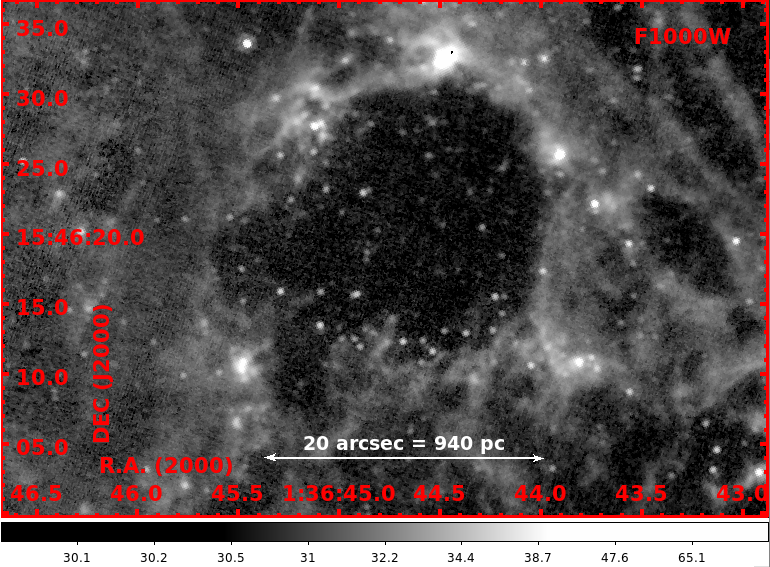}
\includegraphics[width=0.48\linewidth]{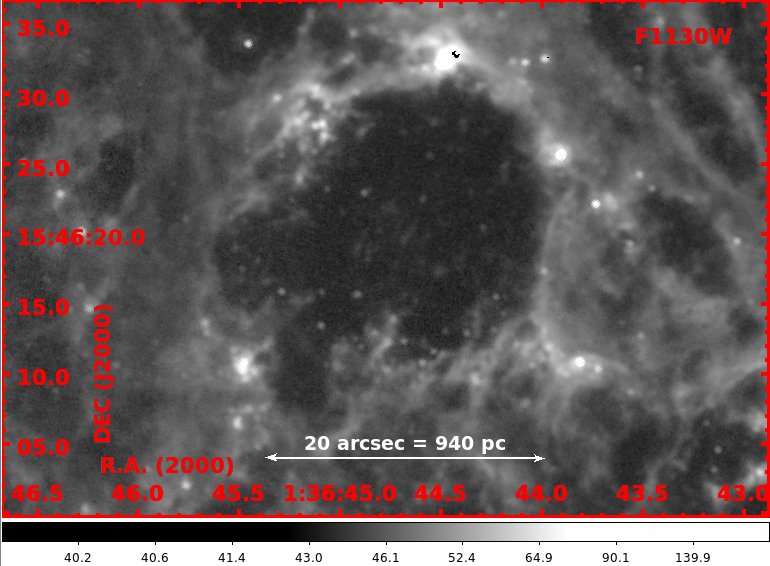}
\includegraphics[width=0.48\linewidth]{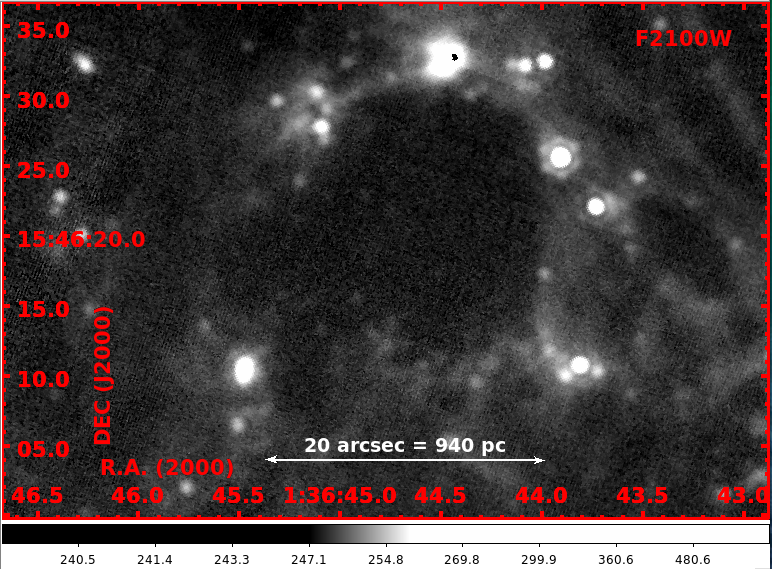}
\par\end{centering}
\caption{
Enlarged view of the bubble in the JWST/{\sc miri} filters used for observation of NGC\,628. Filter names, as defined in Table~\ref{tab:data} are indicated in each panel. All these filters intercept features related to PAH and/or silicates, in addition to continuum radiation present at these wavelengths. The image resolution can be inferred by the size of the point sources in each of these images. The bubble is best traced in F770W and F1130W filters. \cred{A gray-scale bar is given at the bottom of each image, where the numbers are in units of MJy/sr.}
}
\label{fig:bubble_jwst}
\end{figure*}

In Figure~\ref{fig:bubble} we show the F770W image of NGC\,628 from the JWST mission. The image is spectacular for the presence of numerous holes (or bubbles), giving rise to the use of a popular name ``phantom galaxy'' for galaxies with such an appearance. 
\cred{Most of these holes are physically coherent structures where the gas has been evacuated by stellar feedback processes. However, some large holes could be structures between spiral arm suprs \citep[e.g.][]{Williams2022}}.
The largest of these bubbles is marked by a circle of 20 arcsec diameter, which corresponds to a physical size of 940~parsec at the distance of NGC\,628. Interstellar structures of such large dimensions are normally reported in the \hi\ images as \hi\ holes \citep[e.g.][]{Pokhrel2020}. 
Large bubbles can also be formed by the collective action of stellar winds and multiple supernovae \citep{MacLow1988}, in which case they are referred to as superbubbles. The analysis carried out in this work clearly shows the presence of recently formed stars inside the hole that were part of a few dissolved clusters, and hence we refer to this hole as a superbubble.

In this work, we carryout a multiband analysis of this kiloparsec-scale superbubble using the dataset listed in Tables~\ref{tab:data} and \ref{tab:datacube}.
The dataset includes stellar, as well as interstellar medium tracers, offered by the JWST and HST missions. In addition, 
we use the MUSE datacube observed within the PHANGS survey \citep{2018ApJ...863L..21K} to generate an \ha\ map to trace the morphology of the bubble in the ionized gas, 
the CO (2-1) line map from ALMA-PHANGS \citep{2021ApJS..255...19L,2021ApJS..257...43L} to trace the molecular gas, and HI 21~cm line map from VLA-THINGS \citep{2008AJ....136.2563W}
 to trace the atomic gas.
In Figure~\ref{fig:bubble_jwst}, we show an enlarged view of the bubble in four of the MIRI bands. The bubble is traced in all the four bands, but is best defined in the F770W and F1130W filters, which trace the PAH molecules. At the longest JWST wavelength (21~$\mu$m), compact sources, probably embedded stars, can be seen coinciding with the bubble boundary. 

\begin{figure*}
\begin{centering}
\includegraphics[width=0.48\linewidth]{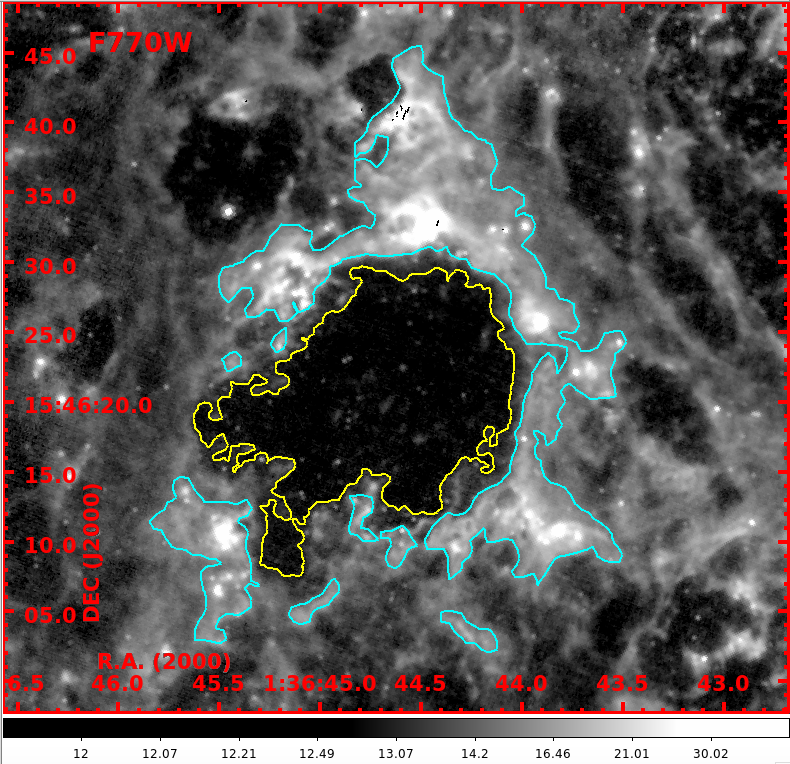}
\includegraphics[width=0.48\linewidth]{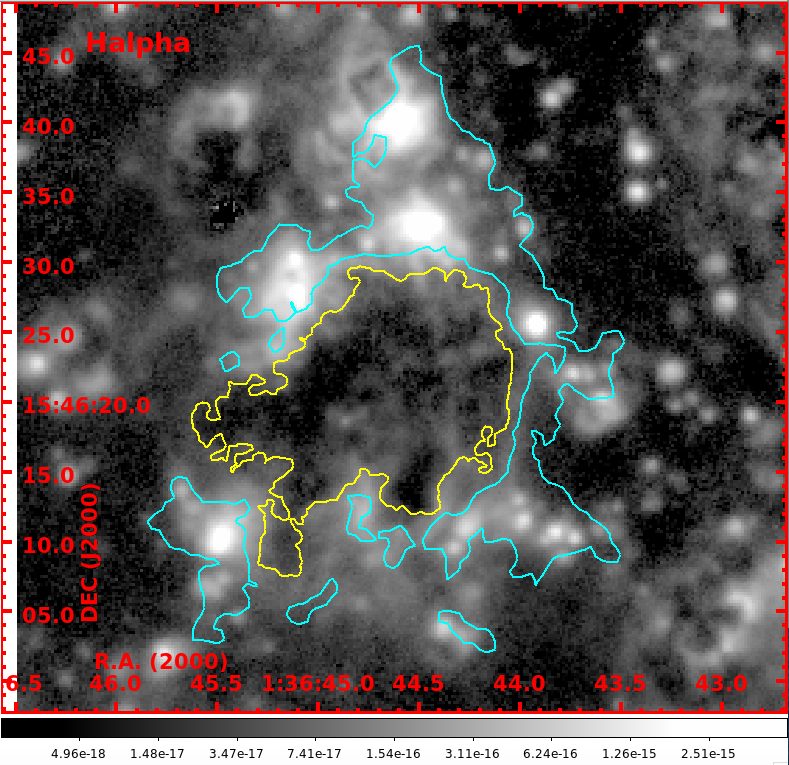}
\includegraphics[width=0.48\linewidth]{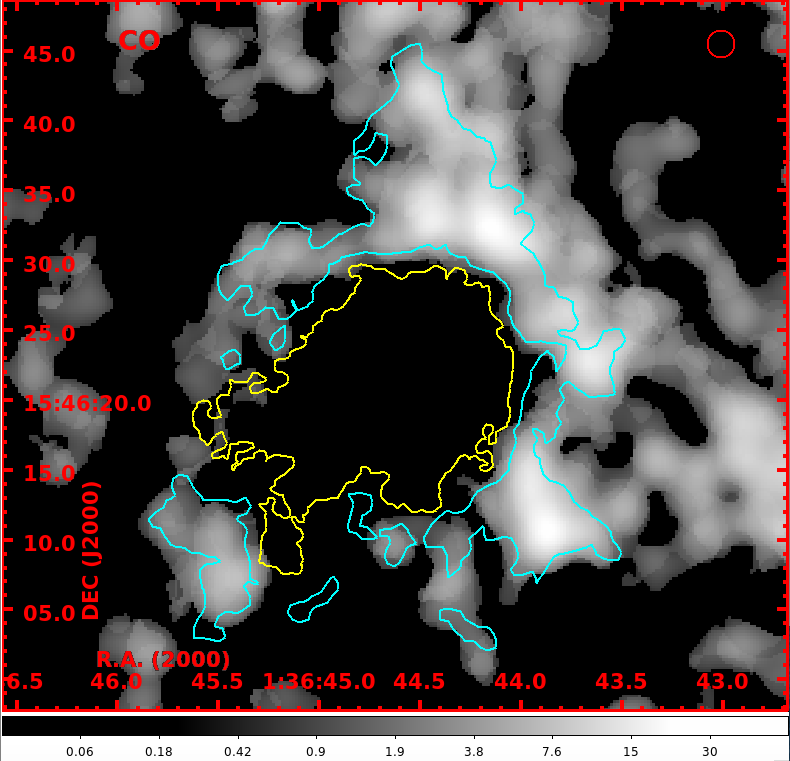}
\includegraphics[width=0.48\linewidth]{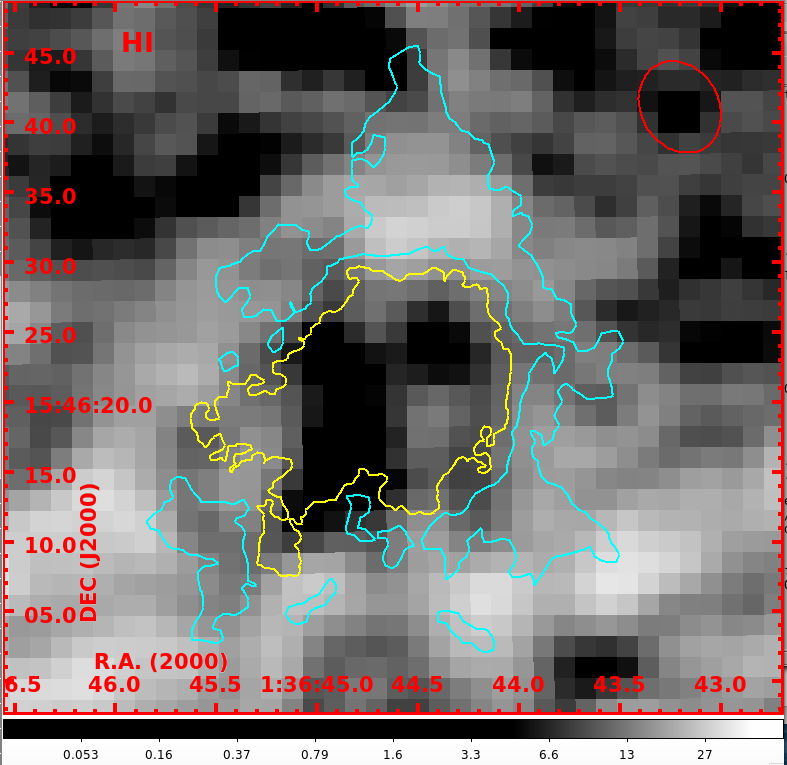}
\par\end{centering}
\caption{
Multiphase view of the bubble. (Top-left) PAH and dust as traced in the JWST F770W filter, (top-right) the ionized gas as traced by the \ha\ nebular line using the MUSE datacube, (bottom-left) the molecular gas as traced by the CO 2-1 line using the ALMA data, and (bottom-right) the neutral atomic gas traced by the 21~cm \hi\ line using the VLA. The beams of the CO and \hi\ images are shown by red circle and ellipses respectively. Contours in all images delineate the brightest (cyan) and the inner boundary (yellow) of the shell, as seen in the F770W image. 
\cred{A gray-scale bar is given at the bottom of each image, where the numbers are in units of MJy/sr, \ergsarc, K\,\kms/pixel and Jy\,m\,s$^{-1}$/beam for the F770W, \ha\, CO and \hi\ images, respectively. The yellow and cyan contours correspond to 12.5 and 15 MJy/sr flux levels on the F770W image, respectively.}
\label{fig:bubble_multiphase}
}
\label{fig:bubble_multiphase}
\end{figure*}

\begin{figure*}
\begin{centering}
\includegraphics[width=0.80\linewidth]{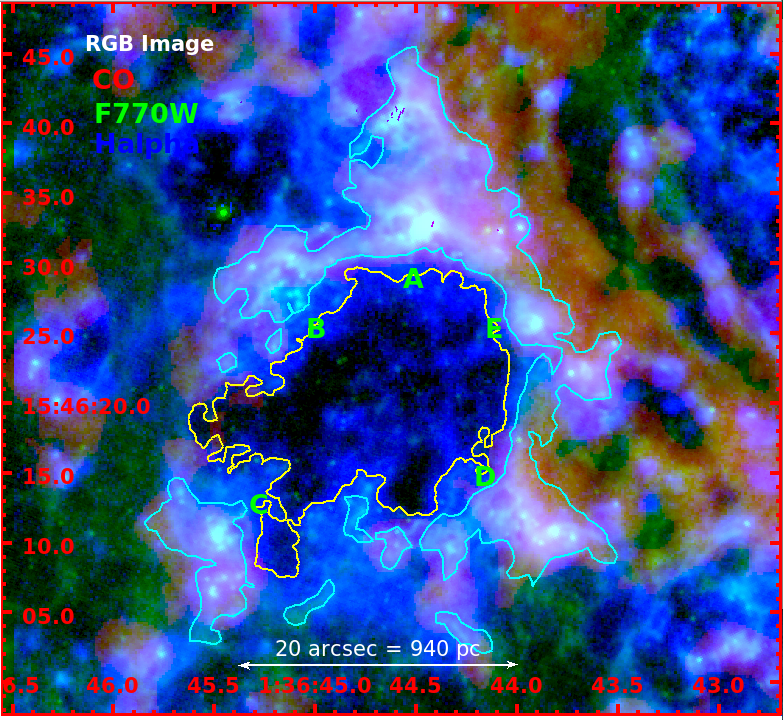}
\par\end{centering}
\caption{
RGB image formed from CO, F770W and \ha\ images as red, green and blue components, illustrating the multi-phase morphology of the bubble. The brightest knots seen in the \ha\ images are marked by letters A to E. The contours are the same as in Figure~\ref{fig:bubble_multiphase}.
}
\label{fig:bubble_multiphase_RGB}
\end{figure*}

NGC\,628 is one of sample galaxies in the PHANGS survey \citep[][]{Emsellem2022} and is observed by the JWST under the title ``A JWST-HST-VLT/MUSE-ALMA Treasury of Star Formation in Nearby Galaxies'' (Proposal ID: 2107; PI: Janice Lee). The JWST is used to carry out imaging observation of the galaxy in both NIRCAM and MIRI cameras. In Tables~\ref{tab:data} and \ref{tab:datacube} we list the details of the publicly available images in the JWST as well as the HST, VLT, ALMA and VLA databases.

\section{Physical characteristics of the bubble}

\subsection{Multi-phase morphology of the bubble}

In Figure~\ref{fig:bubble_multiphase}, we compare the bubble morphology seen in the F770W map with ionized (\ha), molecular (CO), and atomic (HI) gas tracers. We show two contours that delineate the inner and outer boundaries of the bubble in the F770W band and superpose these contours on the rest of the images to facilitate comparison of the multi-phase morphologies. These contours \cred{can be} approximated by circles of inner and outer radii of 8.5 and 13.8~arcsec, which correspond to 400 and 650~pc, respectively. 
The morphology of the ionized gas very closely follows that seen in the F770W filter. The bubble is the brightest between PA=$-30^\circ$ to $150^\circ$  (north, north-west and west), with a well-defined circular-shaped boundary. On the other hand, the inner boundary of the bubble in the diagonally opposite direction (south-east) is corrugated and elongated radially. The western part of the bubble is brighter than the south-east segment in the CO emission also. At the relatively poor spatial-resolution of the \hi\ image, the azimuthal emission is uniform.

The bubble is shown as an RGB image in Figure~\ref{fig:bubble_multiphase_RGB}, where we superpose CO, F770W and \ha\ images as red, green and blue components, respectively. The image facilitates to visualize spatial correspondences and offsets between the molecular, PAH and ionized gas emission. There is bright emission in the shell (delineated by the cyan contours) in all the three tracers. The brightest knots in the \ha, which are identified by letters A to E, are bright in the F770W image also. However, there are noticeable offsets in the emission in the three tracers at different azimuthal locations of the shell. For example,  the CO emission in the bright north-western part is systematically offset outwards. There is a faint diffuse component in \ha\ emission inside the bubble, with no such emission detected in any other ISM tracer. The inner edge of the shell (the thin region between the yellow and cyan contours) is filled with a narrow \ha\ shell, with the PAH lying systematically outside this thin shell traced in \ha.
 
The observed non-circular morphology can be qualitatively understood as effects of an expanding bubble in a non-uniform medium. The bubble center lies in the inter-arm region, with the spiral arm present to the north-west (see Figure~\ref{fig:bubble}), and relatively diffuse gas on the southeast. Hence, the expanding bubble is bounded by the gas in the spiral arm, forming a \cred{semi-circular arc } to the north-west, whereas it is expanding 
\cred{into relatively rarer medium } in the opposite direction, which is the reason for the observed elongation in the direction perpendicular to the spiral arm. 
\cred{The bubble walls are intact even in this direction as evidenced by the almost uniform \hi\ gas surrounding the yellow contours. We discuss the likely 3-D morphology of the bubble in Sec.5.1}.

\begin{figure}
\begin{centering}
\includegraphics[width=0.99\linewidth]{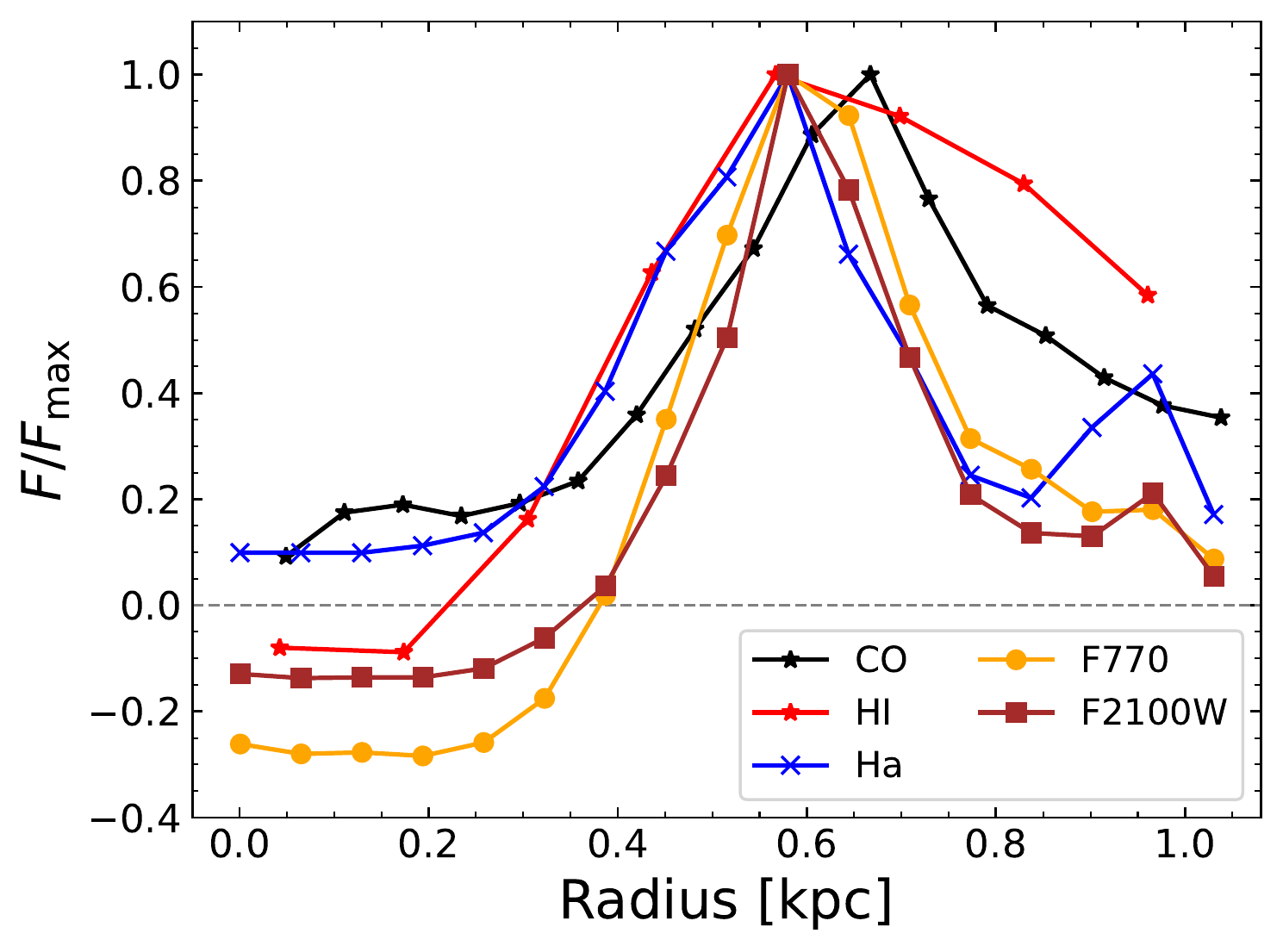}
\par\end{centering}
\caption{
Azimuthally averaged radial intensity profiles of the bubble in the four tracers of the multi-phase ISM (molecular, atomic, ionized and PAH), and the F2100W filter, which besides tracing the PAH emission, identifies the location of on-going star-formation through the hot dust emission. All profiles are normalized to their peak values, which lie close to the inferred bubble radius of 550~pc. See text for details.
}
\label{fig:bubble_rad}
\end{figure}

\begin{figure*}
\begin{centering}
\includegraphics[width=0.75\linewidth]{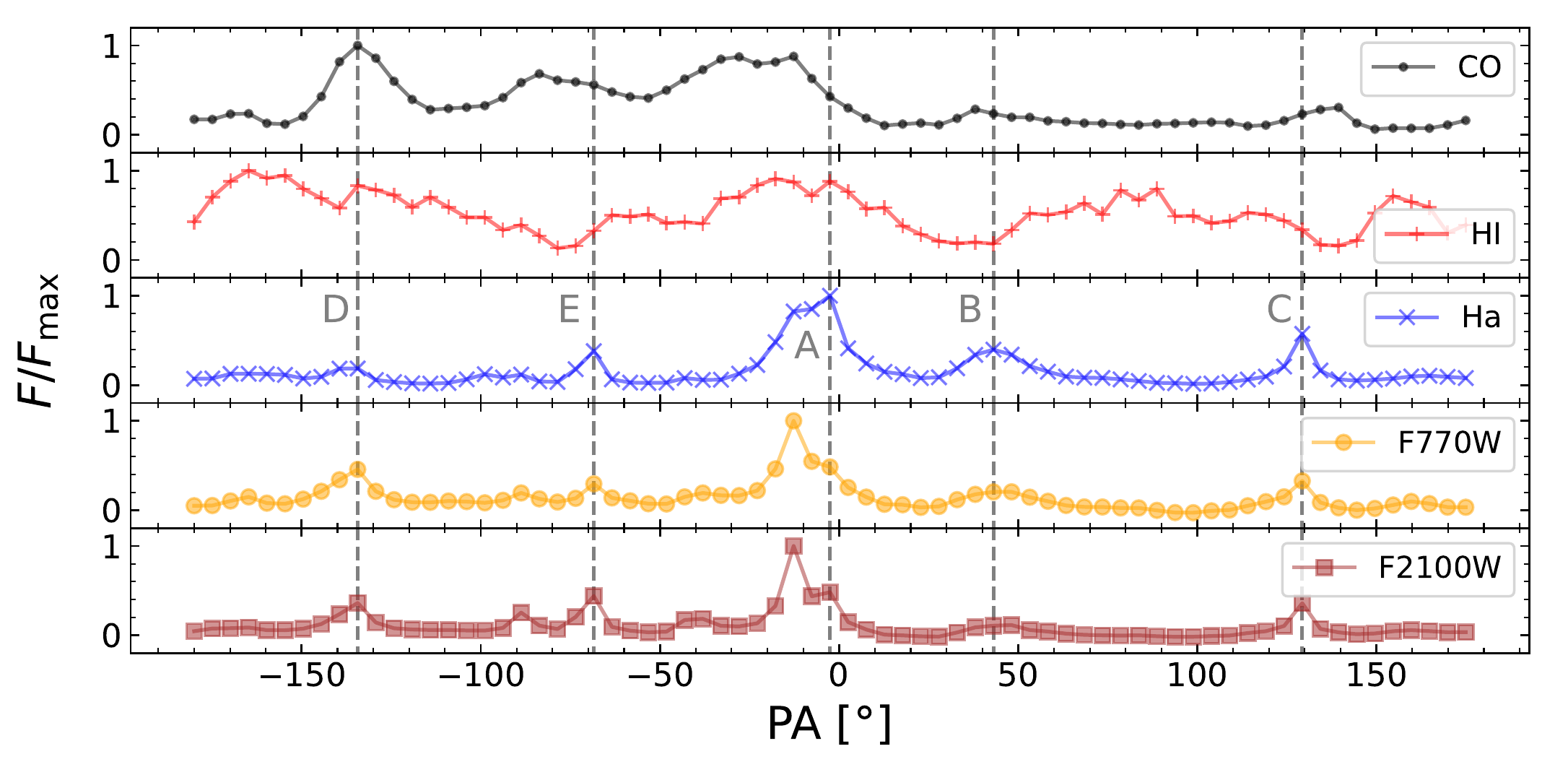}
\par\end{centering}
\caption{
Azimuthal profiles of the bubble where the structures in the radial range of 460 and 930~pc have been averaged. The locations of peaks in the \ha\ profile  are indicated by the vertical lines, and are identified by the letters A to E. There is good correspondence between the peaks in \ha\ and that in F770W and F2100W filters.
}
\label{fig:bubble_pa}
\end{figure*}

\subsection{Radial and azimuthal structure of the bubble}

In Figure~\ref{fig:bubble_rad}, we show the azimuthally averaged radial intensity profiles of the bubble in the four chosen ISM tracers, as well as in the F2100W filter. The latter filter centered at 21~$\mu$m traces active star-forming locations through the continuum emission from hot dust. These profiles were obtained using concentric circular apertures
\cred{on the disk background-subtracted images, where the average value in an annular zone between 1.1 and 1.4~kpc is taken as the disk background. It may be noted that the disk background is the major contribution to the observed flux in \hi, F2100W and F700W images, whereas it is negligible in \ha\ and CO images.}
In all the five plotted tracers, the profile has minimum and maximum intensity at the center of the bubble and the shell, respectively, with the peak emission occurring at a radius=0.55~kpc in \ha, \cred{\hi,} F770W and F2100W filters, and at 0.65~kpc in the CO line. 
\cred{The decrease of intensity outside the shell is the shallowest in \hi, which is the typical characteristic of an \hi\ hole. }
In comparison to \hi, the three other tracers of neutral ISM (CO, F770W, F2100W) show a clear intensity enhancement at the shell, in addition to a dip at the bubble center with respect to the surrounding disk, which are clear characteristics of an expanding shell. The \ha\ profile illustrates that the bubble center has detectable diffuse ionized gas (DIG), with a strength that is comparable in regions outside the shell. The normalized intensities in the two JWST filters inside the bubble are negative, which suggests that the PAH molecules are destroyed inside the bubble.

\begin{figure}
\begin{centering}
\includegraphics[width=0.99\linewidth]{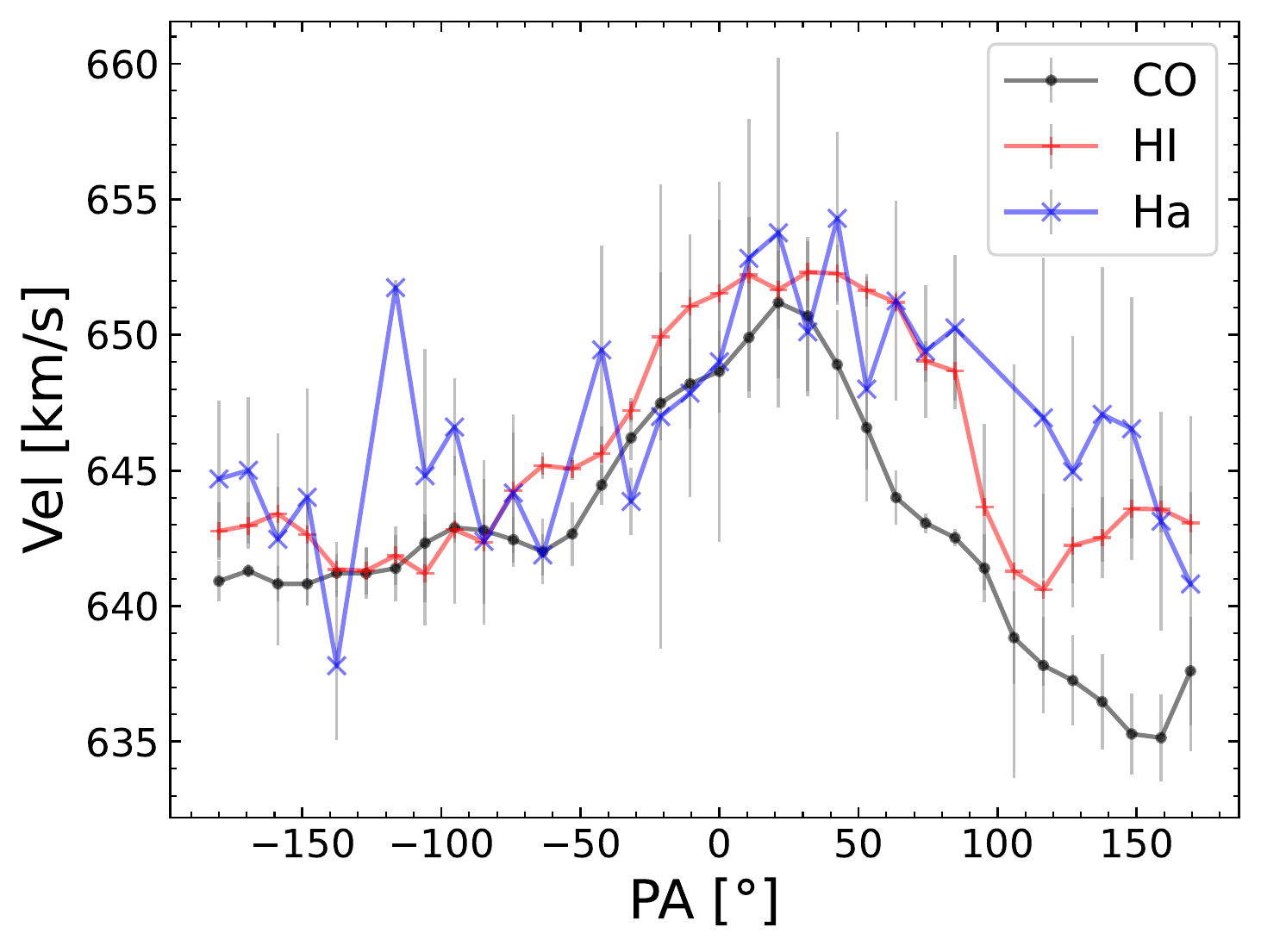}
\includegraphics[width=0.99\linewidth]{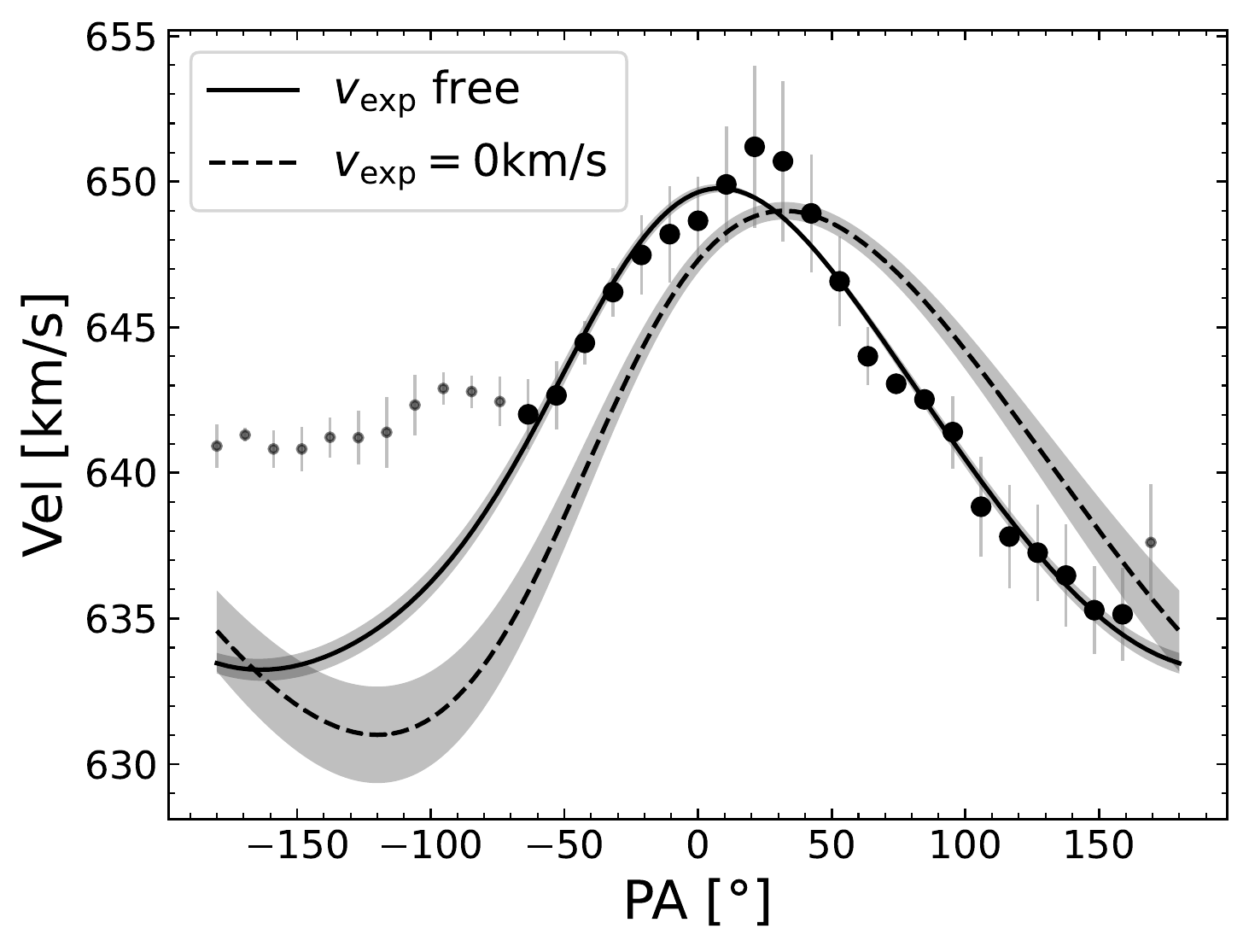}
\par\end{centering}
\caption{
(top) Observed velocities in CO, \hi\ and \ha, along with their error bars, as a function of position angle.
The CO velocities (bottom), which have the best combination of the spatial and spectral resolution, are fitted with an expanding bubble model in the PA range of $-60^\circ$ to 160$^\circ$. The model with expansion (solid curve) fits these data better than a model without expansion (dashed curve). Neither of these fits can reproduce the observed 
almost constant velocity in the  PA=$-$70$^\circ$ to $-$180$^\circ$ range. See text for details. 
}
\label{fig:bubble_velocity}
\end{figure}

It can be inferred from images presented in Figure~\ref{fig:bubble_multiphase} that the shell shows considerable azimuthal structures. We show the azimuthal intensity profiles in the five chosen tracers in Figure~\ref{fig:bubble_pa} which allows us to compare the observed morphology in different phases of the ISM. All emission in an annular zone between radii 0.46~kpc to 0.93~kpc are summed to obtain these profiles. The profiles show narrow peaks superposed on a slowly varying extended emission. 
The extended emission comes from diffuse gas, which is expected to be predominantly the gas that is swept-up as the bubble expanded. On the other hand, the peaks correspond to compact knots present in the shell. The contrast between the extended emission and the compact knots is the highest in the \ha\ profile, where we can identify five zones of emission, which we identify in the figure by letters A to E. Each of these five peaks in \ha\ has a corresponding peak in F770W, and F2100W profiles. Examination of the image in these two JWST filters reveals that the compact sources dominate to the observed emission at these peaks rather than the diffuse component from PAH. The emission from compact sources at MIR wavelengths originates from the hot dust surrounding the recently formed stars. These knots are indeed new star-forming sites triggered by the positive feedback of the expanding bubble \citep{2019ApJ...870...32K}. We will address this issue later in the discussion section.

\subsection{Kinematics of the gas in the shell}

We used the {\rm velocity maps} in CO, \hi\ and \ha\ to obtain kinematical information of the shell gas in molecular, atomic and ionized phases. We illustrate the results in Figure~\ref{fig:bubble_velocity}, where we plot the observed  velocities against the position angle, PA. In the top panel, we show the velocities obtained in the three tracers. A systematic variation of velocity as a function of PA is seen in all the three tracers. A uniformly expanding ring is expected to follow sinusoidal curve in this plot. However, given the large size of the bubble, the contribution of the galactic rotation to the observed velocity profile cannot be ignored. We hence fitted the observed
velocities $V_{\rm{obs}}$ with a sinusoidal model considering the expansion centered in the bubble, plus a rotation around the center of the galaxy, and obtained the expansion velocity $V_{\rm exp}$ using the following equation \citep{1978ARA&A..16..103V}:
\begin{equation}
V_{\rm{obs}}=V_{\rm{sys}}+V_{\rm exp} \sin (i) \sin (\theta)+V_{\rm rot} \sin (i) \cos (\theta^{\prime})
\label{eq:kinmod}
\end{equation}
where $i$ is the inclination angle of the disk, $\theta$ is the azimuthal angle in the plane of the galaxy centered in the bubble, which is related to the measured position angle $\rm{PA}$ through the equation $\tan(\theta)=\tan(\rm{PA}-\rm{PA}_0)/\cos(i)$, where $\rm{PA}_0$ is the position angle of the major axis. $\theta^{\prime}$ is the azimuthal angle in the plane of the galaxy centered in the galaxy since the rotation is around that center, and it is equally related to the position angle centered in the galaxy, PA$^{\prime}$. We can obtain the position angle centered in the galaxy, PA$^{\prime}$, in terms of the position angle centered in the bubble PA by solving:
\begin{equation}
\tan(\rm{PA^{\prime}})=-\frac{\rm{\Delta} \alpha_{0}-r\sin{\rm{PA}}}{\rm{\Delta} \delta_{0}+r\cos{\rm{PA}}}
\label{eq:pa_gal}
\end{equation}
where $r=$650~pc is the radius of the bubble estimated from the azimuthal averaged CO intensity profile (black line in Fig. \ref{fig:bubble_pa}).   $\rm{\Delta} \alpha_{0}$ and $\rm{\Delta} \delta_{0}$ are the sky coordinates difference between the center of the bubble, and the center of the galaxy: $(\rm{\Delta} \alpha_{0},\rm{\Delta} \delta_{0})=(\alpha_{\rm{bubble}}^{\rm{center}},\delta_{\rm{bubble}}^{\rm{center}})-(\alpha_{\rm{gal}}^{\rm{center}},\delta_{\rm{gal}}^{\rm{center}})$. We have fixed $V_{\rm{sys}}=653\rm{km/s}$ \citep{1993ApJS...88..383L}, $\rm{PA_0}=26.4^\circ$; $i=21.5^\circ$ \citep{2006MNRAS.367..469D}, $(\alpha_{\rm{gal}}^{\rm{center}},\delta_{\rm{gal}}^{\rm{center}})=(01\rm{h}36\rm{m}41.747\rm{s},15\rm{d}47\rm{m}01.18\rm{s})$, and $(\alpha_{\rm{bubble}}^{\rm{center}},\delta_{\rm{bubble}}^{\rm{center}})=(1\rm{h}36\rm{m}44.6172\rm{s},15\rm{d}46\rm{m}21.4\rm{s})$.

We have fitted the sinusoidal model from Eq.~\ref{eq:kinmod} in terms of PA. We fitted the observed velocities using the CO data since it offered us the best combination of spatial and kinematical resolutions among the available data. The observed data follows a sinusoidal curve in the PA range of $-60^{\circ}$ to $160^{\circ}$,
where the maximum and minimum values of observed velocities are clearly visible. The observed velocities outside this PA range are almost approximately constant, which would require a model more complex than the model we are fitting. This part where velocity is almost constant corresponds to the western arc where the bubble grazes the spiral arm. The encounter of the swept-up material with the spiral arm is expected to  decelerate the expansion of the bubble, which is probably the reason for the almost constant velocity in this part of the bubble. The results of the fits performed for PA$=-60^{\circ}$ to PA$=160^{\circ}$ (darker points) are shown in the bottom panel of Fig.~\ref{fig:bubble_velocity}. We show two fits, one where we force the expansion velocity, $V_{\rm{exp}}$, to be zero (dashed line) and the other where we let it free (solid line). It can be clearly seen the fit does not replicate the observed velocities  when not considering expansion. 

From these data, we determine $V_{\rm exp}=-$11.9$\pm$0.7\,\kms, and $V_{\rm rot}=108\pm$2\,\kms. The negative value of the expansion means that the approaching side of the expansion is in the positive range of PA, the eastern side of the bubble. This fact is consistent with the trailing spiral arms in this galaxy. 

\subsection{Mass of the gas in the shell}

Most of the gas in the shell is in molecular form. Using a CO(2-1) to H$_2$ conversion factor $\alpha_{\rm{CO}}=6.1\thinspace(\rm{K\thinspace km/s})^{-1}$ corresponding to the Galactic CO(1-0) value \citep{2013ARA&A..51..207B} and a CO(2-1)/CO(1-0) ratio of 0.7 \citep{2009AJ....137.4670L}, we determine a total molecular hydrogen mass of 1.52$\times10^7$~\msol\ in the shell between the 400 to 700~pc annular ring. The atomic mass in the same zone is 5.4$\times10^6$~\msol, giving a total gas mass of 2.06$\times10^7$~\msol. The CO and \hi\ intensities per pixel is below the 3-$\sigma$ detection limit  in pixels inside the bubble. We estimate an upper limit of $6\times10^6$~\msol\ inside a radius of 460~pc.

\begin{figure*}
\begin{centering}
\includegraphics[width=0.32\linewidth]{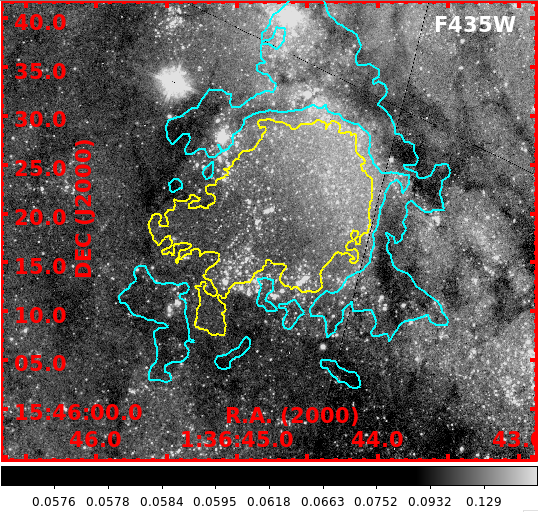}
\includegraphics[width=0.32\linewidth]{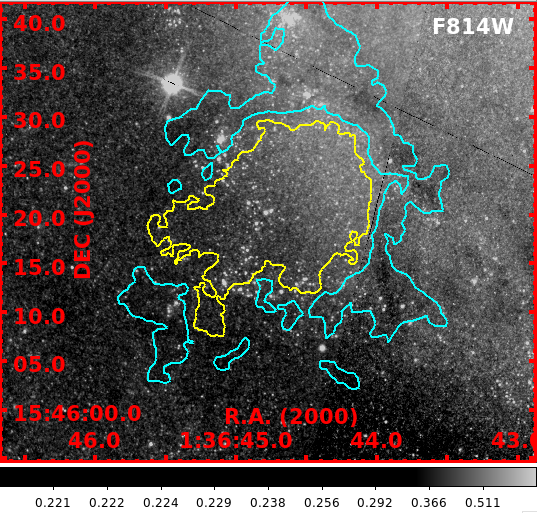}
\includegraphics[width=0.32\linewidth]{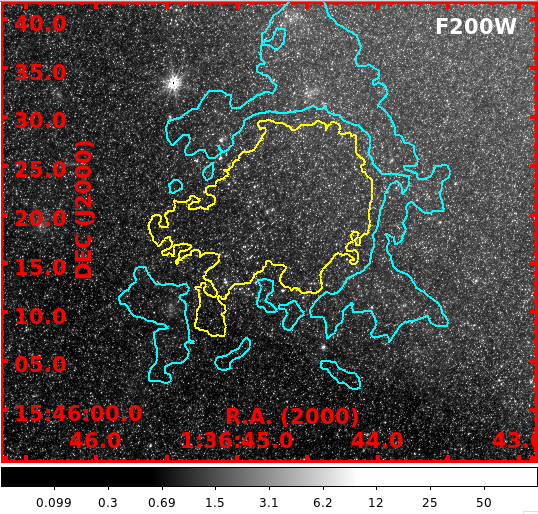}
\includegraphics[width=0.32\linewidth]{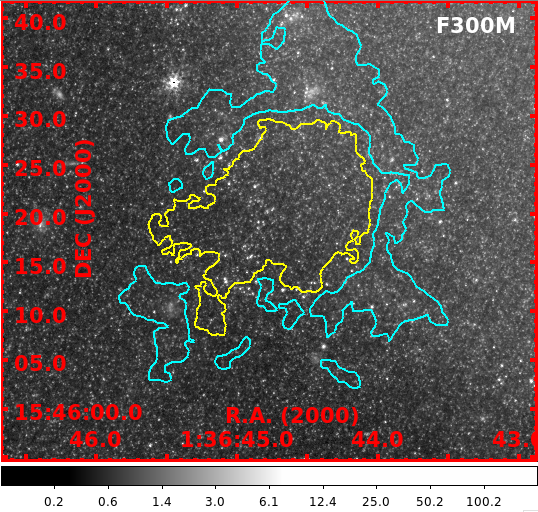}
\includegraphics[width=0.32\linewidth]{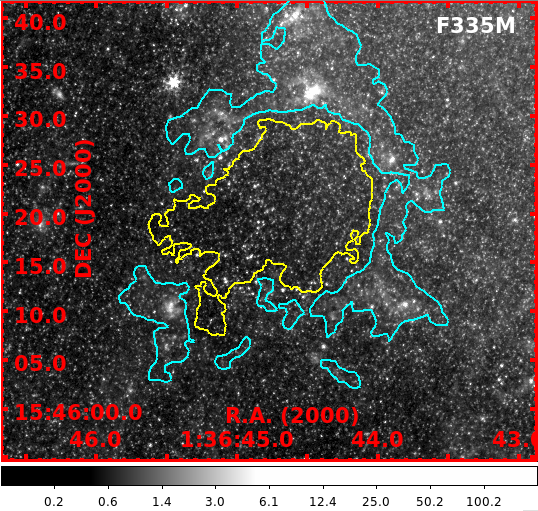}
\includegraphics[width=0.32\linewidth]{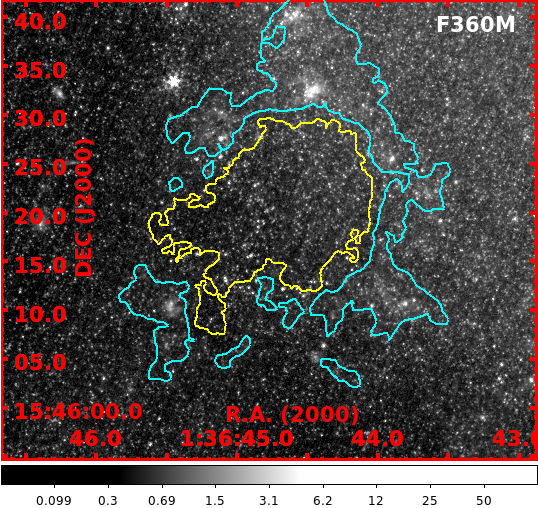}
\par\end{centering}
\caption{
Enlarged view of the bubble in filters that trace stellar continuum from 0.43~$\mu$m to 3.6~$\mu$m. Note the bright stars among the resolved population in the F435W image, whose prominence decreases at longer wavelengths.  The cyan contours delineate the brightest part of the bubble in F770W filter; the blue contours trace the inner boundary of the bubble.
\cred{A gray-scale bar is given at the bottom of each image, where the numbers are in units of electron/s for the F435W and F814W images and MJy/sr  for rest of the images.}
}
\label{fig:bubble_stars}
\end{figure*}

\begin{figure*}
\begin{centering}
\includegraphics[width=0.49\linewidth]{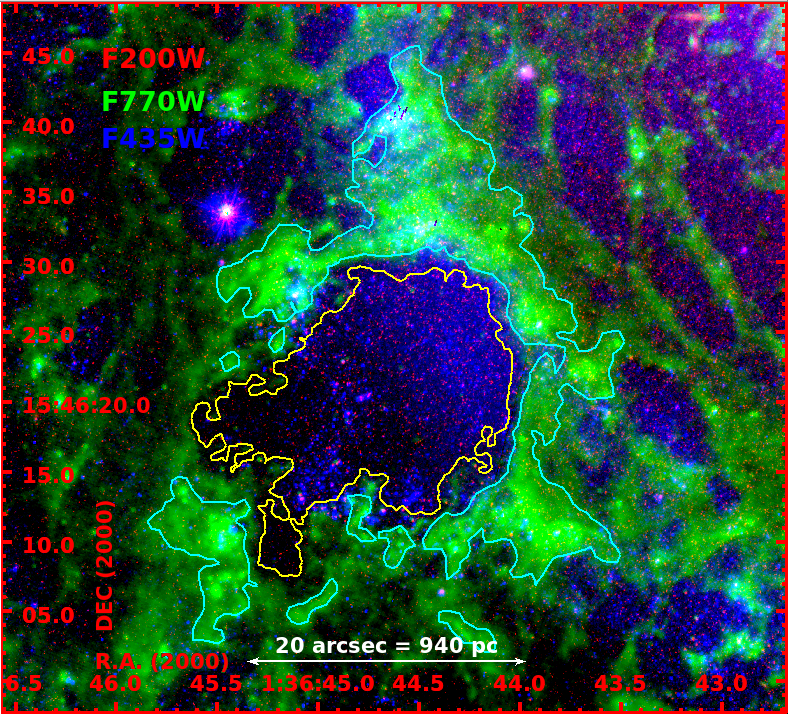}
\includegraphics[width=0.49\linewidth]{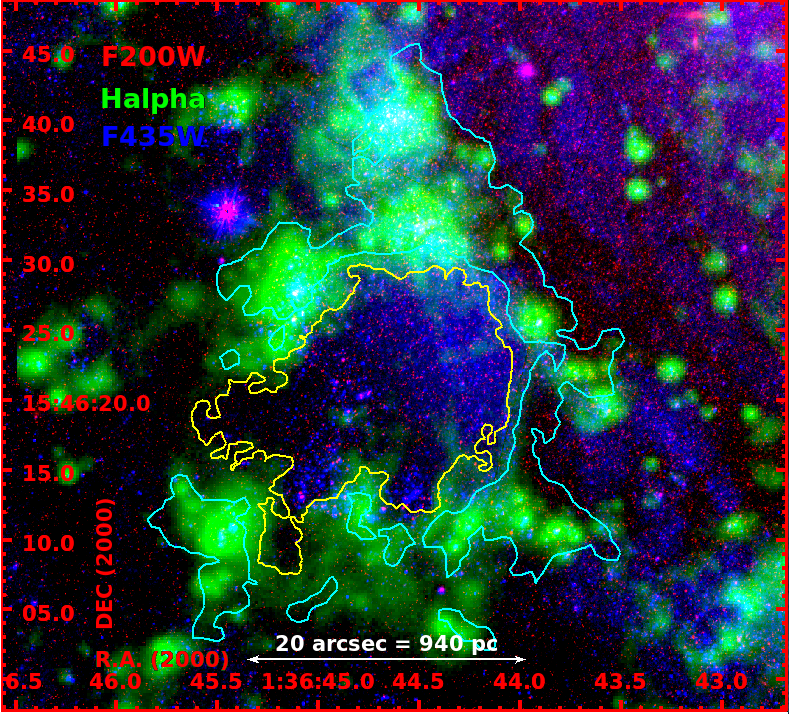}
\par\end{centering}
\caption{
RGB images to illustrate the confinement of bluer stars to the inner boundary of the bubble. In comparison redder stars are more uniformly distributed inside and outside the bubble. The blue and red components of both the left and right images are the same (shown in corresponding colours on the top-left corners), whereas the F770W image is used on the left and \ha\ image on the right. The F770W contours shown in the previous figures are superposed.  
}
\label{fig:bubble_stars_RGB}
\end{figure*}

\section{Stellar populations inside the bubble}

A superbubble is created due to the mechanical power deposited to the interstellar medium from winds from massive stars and multiple supernova explosions. 
Superbubbles are short-lived structures and hence the relatively lower-mass stars of the stellar population that contained these massive stars should still be present inside the bubble. Stars need to be more massive than 8~\msol\ to explode as supernova, which implies that there should be stars younger than 40~Myr inside the bubble.
On the other hand, the stellar population could be as young as 4~Myr if it contained stars as massive as 100~\msol\ and the bubble got created in one of the first explosions of supernova in the cluster. In the STARBURST99 models, a Simple Stellar Population (SSP) produces its maximum mechanical power between 4 to 20~Myr.
We used the HST/ACS and JWST/NIRCam images of NGC\,628 to trace and characterize the stellar population that might be responsible for the creation of the superbubble.

Magnitude at the turn-off point of 8~\msol\ star is F814W=24~mag, which is above the detection limit of the HST observations and hence post-main sequence stars of clusters as old as 40~Myr can be detected. For younger clusters, stars in the upper end of the main sequence can also be detected.

In Figure~\ref{fig:bubble_stars}, we display the images in the F435W and F814W filters of HST and F200W, F300W, F335W and F360W of the JWST/NIRCam filters. The HST images needed a small astrometric correction which we carried out by visually identifying stars common in the F200W JWST image. We superpose the two F770W contours defined in Figure~\ref{fig:bubble_multiphase} in all the images which enables us to identify the location of stars with respect to the bubble. In the HST images, we clearly observe an overdensity of stars inside the bubble. 
In Figure~\ref{fig:bubble_stars_RGB} we show RGB images, where the HST/F435W and JWST/F200W images are chosen as the blue and red components, respectively. For the green component, we choose JWST/F770W image in the left and MUSE/\ha\ image on the right. The  figures illustrate that the blue stars are selectively confined inside the bubble, whereas the red stars are almost uniformly distributed inside and outside the bubble. The blue stars are in general brighter and are detected in both the HST and JWST filters, whereas the majority of the JWST-detected stars are mostly undetected in the HST/F435W filter. We carry out aperture photometry of all stars inside the bubble to determine the age of the blue population stars.

\subsection{Detection and photometry of stars in the bubble}\label{sec:Sextractor}

We used the SExtractor \citep{Bertin1996} code to make a catalogue of stars detected in the HST and JWST/NIRCam images. Given the different spatial resolutions and pixel scales of the images used here (see Table~\ref{tab:data}), we ran SExtractor independently on each image and then obtained a list of common stars in any two images based on the RA and DEC coordinate matches. Stars in the two catalogues whose coordinates differ by less than 0.15~arcsec are considered counterparts of each other. For each catalogued source, we saved the FWHM, the AREA above the 4-$\sigma$ threshold, and magnitudes in concentric apertures of multiple radii. We used the zeropoints given in Table~\ref{tab:data} to convert the image counts to apparent magnitudes in the VEGA system.

\begin{figure}
\begin{centering}
\includegraphics[width=0.95\linewidth]{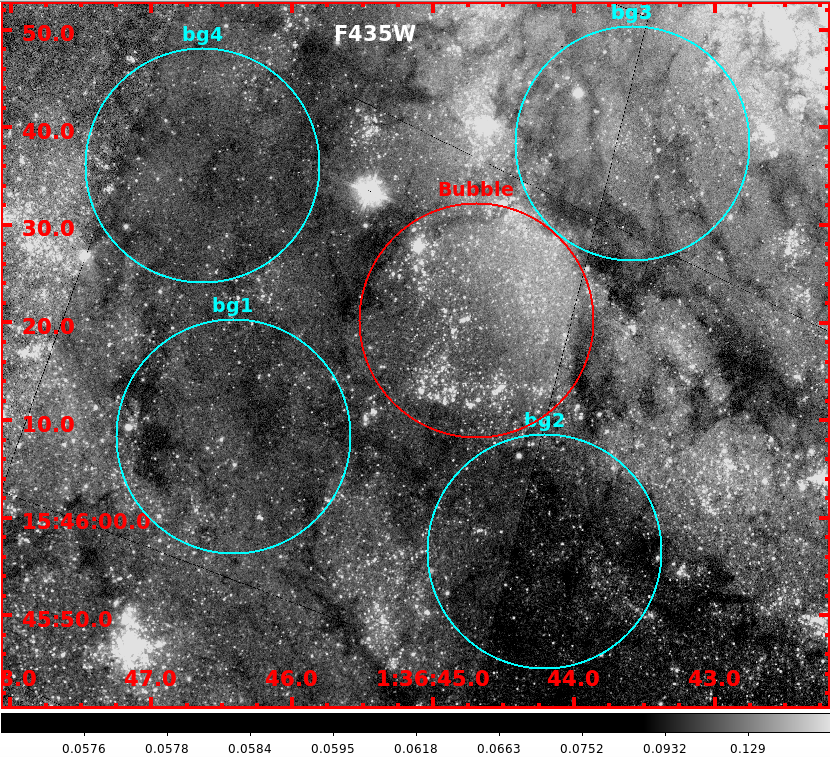}
\par\end{centering}
\caption{
An image from HST/F435W showing the location of the bubble detected in JWST bands\cred{, where the gray-scale bar units are electron/s}. Other four regions around the cavity,  (bg1, bg2, bg3 and bg4) are indicated by cyan circles with the same diameters. These regions were selected to be representative of the stellar population in the disk and compare with the stars inside the bubble. See text for details.
}
\label{fig:bg_areas}
\end{figure}

We defined a circular area of 450~pc radius centered on the bubble to create lists of bubble stars in each filter. This list contains stars physically inside the bubble as well as the disk stars in the line of sight. The latter contribution can be statistically determined and subtracted by analysing the color-magnitude properties of the stars in the immediate vicinity of the bubble. The disk brightness varies considerably around the bubble, and hence it is necessary to take into account this variation in order to infer a possible excess population of stars that are physically inside the bubble. We defined four regions around the bubble to sample the typical projected  disk population at the location of the bubble. These regions are shown in Figure~\ref{fig:bg_areas}.

Given that the F200W has the best spatial resolution among all the images, we used the catalogue in this filter as the base catalogue and obtained the JWST colours of these stars. We found that the majority of the JWST stars are not detected in the F435W images and are only marginally detected in the F814W images. We hence made a separate catalogue of stars detected in both the F435W and F814W filters. We use the magnitudes and colours in aperture of 0.15~arcsec radius for constructing colour-magnitude diagrams. The 3-$\sigma$ limiting magnitudes in these filters are given in Table~\ref{tab:data}.

\begin{figure*}
    \centering
    \includegraphics[width=0.45\textwidth]{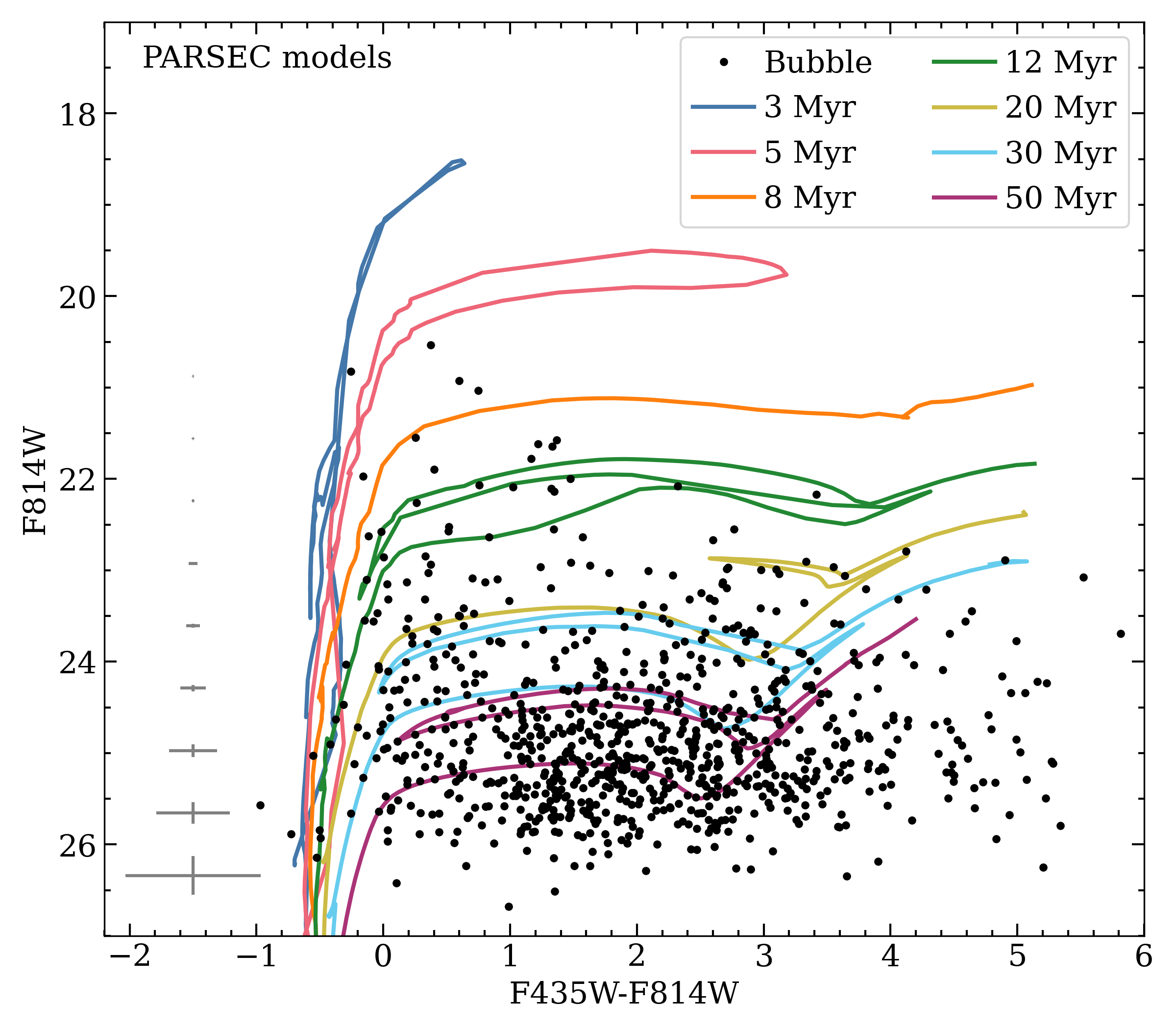}
    \includegraphics[width=0.45\textwidth]{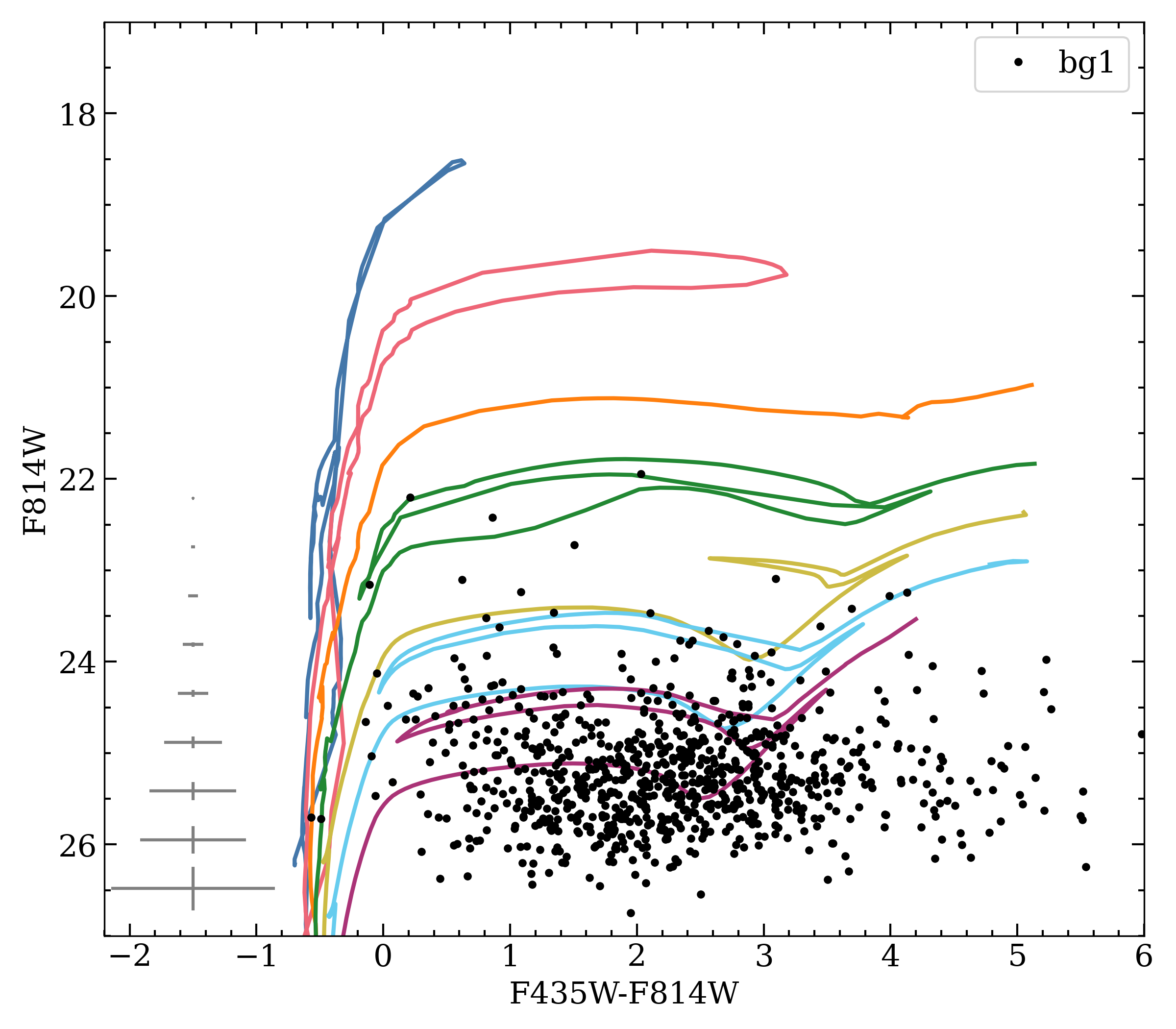}
\caption{CMD of the HST sample of stars inside the bubble (left) and in the selected disk region (bg1; right). Seven isochrones from PARSEC models are superposed with their ages indicated in the legends. Note than most of the disk population contains stars of 50~Myr or older.
}
    \label{fig:cmd_bub}
\end{figure*}

\subsection{Colour-magnitude diagrams of the bubble and disk stars}\label{sec:Sext}

In Figure~\ref{fig:cmd_bub}, we compare the stellar populations in the bubble (left) with that of a typical disk (right) in the immediate vicinity in a CMD using the HST magnitudes in the Vega systems. In both the graphs, we plot isochrones at selected ages using the PAdova and TRieste Stellar Evolution Code \citep[PARSEC;][]{Bressan2012}. A comparison of the locations of stars in these plots reveals that most of the disk population contains stars of 50~Myr or older. A subset of stars in the region bounded by $-0.5\leq F435W-F814W\leq 1.5$ and $20\leq F814W\leq 24$ in the bubble CMD clearly stand out. We suspect that these stars, most of which are younger than 30~Myr, are responsible for the creation of the expanding bubble. We carryout below a quantitative analysis of the stellar populations in the bubble and the disk in order to determine the ages of this relatively younger stellar population in the bubble.

\subsection{Stellar population analysis}

We used the statistical model from \cite{Alzate2021} to compute the ages of the stellar populations in the bubble and disk samples. This approach uses the Bayes theorem in its hierarchical mode to infer the star formation history (SFH) of any resolved stellar population from the photometry of its stars. In this framework, the algorithm computes the probability distribution of the position of stars in the color-magnitude diagram as a linear combination of theoretical isochrones of distinct ages, each with a weight $a_i$, whose numerical value is fixed by the fraction of total number stars fitting a given isochrone \citep{Small2013}. The statistical approach we have followed here has advantages over the traditional isochrone fitting method, as this novel method obtains SFH rather than the best-fit age of the single population.

Each photometric magnitude from SExtractor is modeled as a random sampling from a statistically independent normal distribution function (or the likelihood), where the mean ($\mu$) of the distribution is the predicted magnitude from stellar models and its standard deviation ($\sigma$) is the magnitude error from SExtractor.
The prior Probability Distribution Function (PDF), from which photometric magnitudes and stellar counts are predicted, is obtained using the Initial Mass Function (IMF) and the stellar isochrones, and it is probabilistically  conditioned to $a_i$ through the ${\rm P}(f_{j}^{1}f_{j}^{2}f_{j}^{3}\vert \bm{a})$ term in Eq.~\ref{eq:bayes}.
The product of the likelihood and the prior, properly normalized and computed for the $j^{\rm th}-$star of the sample, gives the probability that the magnitudes of an observed star in different filters were sampled from the prior set of isochrones.
Tables \ref{tab:model_param} and \ref{tab:model_pdf} (see Appendix) show the parameters and the probability distributions entering in the statistical model.
Using all these components and selecting a hyper-prior for $a_{i}$, the posterior PDF  of stellar fractions is computed through

\begin{equation}\label{eq:bayes}
    p(\bm{a} \vert \bm{F}) \propto p(\bm{a}) \prod_{j=1}^{N_{D}} \prod_{k=1}^{3} \int{ \frac{S(F_{j}^{k})\mathcal{L}(F_{j}^{k} \vert \widehat{F}_{j}^{k}) \ p(f_{j}^{1}f_{j}^{2}f_{j}^{3}\vert \bm{a})}{\ell(\bm{a},S)} } d{f_{j}^{k}},
\end{equation}
where subscript $j$ indicates the $j^{\rm th}$ star and the superscript $k$ indicates the photometric magnitudes of the star; $(f^{1},f^{2},f^{3})$=F435W, F555W, F814W. In Eq.\,\ref{eq:bayes}, $p$ and $\mathcal{L}$ stands for PDF  and likelihood function, respectively.
Thereby, $p(\bm{a})$ and $p(f_{j}^{1}f_{j}^{2}f_{j}^{3}\vert \bm{a})$ are the hyper-prior and the prior of the theorem.
The normalization constant ($\ell$) of the posterior PDF will depend on the completeness of the sample.
With an inappropriate normalization, the posterior will predict incorrect stellar counts, and the inference of the SFH will be biased.
We considered two cases to deal with the incompleteness of any sample, computing the posterior PDF for: i) a complete sample ($S(F_{j}^{k})=1$) and ii) a magnitude-limited sample (star counts are complete only for stars brighter than the limiting magnitude, $S(F_{j}^{k})=1$ if and only if 
F555W$\leq$F555W$_{\rm lim}$.
In the second case, counts fainter than the limit will have negligible probability and do not contribute information to the solution.
Once the components of the Bayes theorem are defined and restricted to the completeness intervals, we sampled the marginal posterior distributions of the stellar fractions performing a Montecarlo sampling of $10^{5}$ iterations.
The $a_i$ median estimates the contribution of the  $i$-th isochrone, and the (10\%, 90\%) percentiles delimit the credibility intervals of $a_i$.

We transform these $a_i$ values per interval of age (or per isochrone) to star formation rates (SFRs), assuming a stellar IMF. The number of stars per interval of age is given by
\begin{equation}
    N_{i}=a_i\cdot N_{s},
\end{equation}
where $N_{s}$ is the total number stars of the sample. For a magnitude-limited sample, we obtain $N_i$ as the number of stars brighter than the $F555W_{\rm lim}$, and the number stars fainter than this limit are unknown. Therefore, we calculate  the number of stars below $F555W_{\rm lim}$ by integrating the double power law IMF:
\begin{equation}
\phi(M) = 
\begin{cases}
    C_1\ M^{-\alpha_1}\ \ \ {\rm if}\ \ \ M_l \le M \le M_c\cr
    C_2\ M^{-\alpha_2}\ \ \ {\rm if}\ \ \ M_c \le M \le M_{\rm lim}.
\end{cases}
\label{phi2}
\end{equation}
\noindent as
\begin{equation}\label{eq:N.below}
    N_{\rm lim}(i) = \int_{M_l}^{M_{lim}(i)}\phi(M)\cdot dM,
\end{equation}
\noindent where $M_l=0.1$~\msol\ is the lower mass boundary of the IMF and $M_{\rm lim}(i) = M_{i}(F555W_{\rm lim})$ is the mass of the $i^{\rm th}$-isochrone at the magnitude limit. The IMF in Eq.\,\ref{eq:N.below} is normalized to $N_i$, such that,
\begin{align}
    C_2 &= \frac{ (1-\alpha_2)N_{i} }{ M_{x}^{1-\alpha_2}-M_{c}^{1-\alpha_2} },\nonumber\\
    C_1 &= C_2\cdot M_{c}^{\alpha_1 - \alpha_2}.
\end{align}
We use $M_{c}=0.5$ \msol, slopes $\alpha_1=1.3$ and $\alpha_2=2.3$ according to \citet{Kroupa2001}. Thus, the corrected number of stars is
\begin{equation}\label{eq:ncorr}
    N_{\rm corr}(i) = N_{i} + N_{\rm lim}(i).
\end{equation}
Evolutionary effects are not taken into account in Eq.\,\ref{eq:ncorr}, that is, $N_{\rm corr}$ does not include evolved high mass stars, $[M_{\rm lim}(i),M_u]$, that are evolved off the CMD. Here $M_{u}=350$ \msol\  is the maximum mass that a single star may have in the IMF. However, these stars are fewer and their contribution to the number and the mass of the whole population is negligible.

In this theoretical framework, the expected mean mass of $N_{\rm corr}$ stars is
\begin{equation}
    \langle M \rangle \approx \int_{M_{l}}^{M_{u}}M\cdot\phi(M)\cdot dM,
\end{equation}
\noindent and the star formation rate at the $i^{\rm th}$ age interval is
\begin{equation}\label{eq:sfr}
    SFR_{i} =
    \frac{\langle M\rangle\,N_{\rm corr}(i)}{\Delta t_i} =
    \frac{\langle M\rangle\,N_{\rm corr}(i)}{2.30258\,t_i\,\Delta({\rm log}_{10}\,t_i)},
\end{equation}
where $t_i = 13.7 {\rm Gyr} - {\rm Age}$.

\subsubsection{Isochrones}

We acquired stellar isochrones from version 1.2S of 
PARSEC \citep{Bressan2012}.
This version of the code computes solar-scaled evolutionary tracks for stars with masses from 0.1 \msol to 350 \msol, for metallicities ranging from Z=0.0001 to Z=0.06 \citep{Chen2015}.
We used 23 isochrones\footnote{\url{http://stev.oapd.inaf.it/cgi-bin/cmd}} selected with the following prescription

\begin{itemize}
    \item $Z=Z_{\odot}$,
    \item $\log({\rm Age}/{\rm yr})=[6,8.2]$ dex,
    \item $\Delta\log({\rm Age}/{\rm yr})=0.1$ dex,
    \item $M=[0.1,350]$ M$_{\odot}$.
\end{itemize}

We used PARSEC magnitudes in the HST/ACS photometric system anchored to Vega zero-point reference.
We selected $\log({\rm Age}/{\rm yr})=8.2$ dex as the upper limit of our model grid, motivated by the fact that the turn-off locus of the isochrone at this age is around the limiting magnitude of the bubble and the background samples (F555W=27~mag)


\begin{figure}
    \centering
    \includegraphics[width=0.45\textwidth]{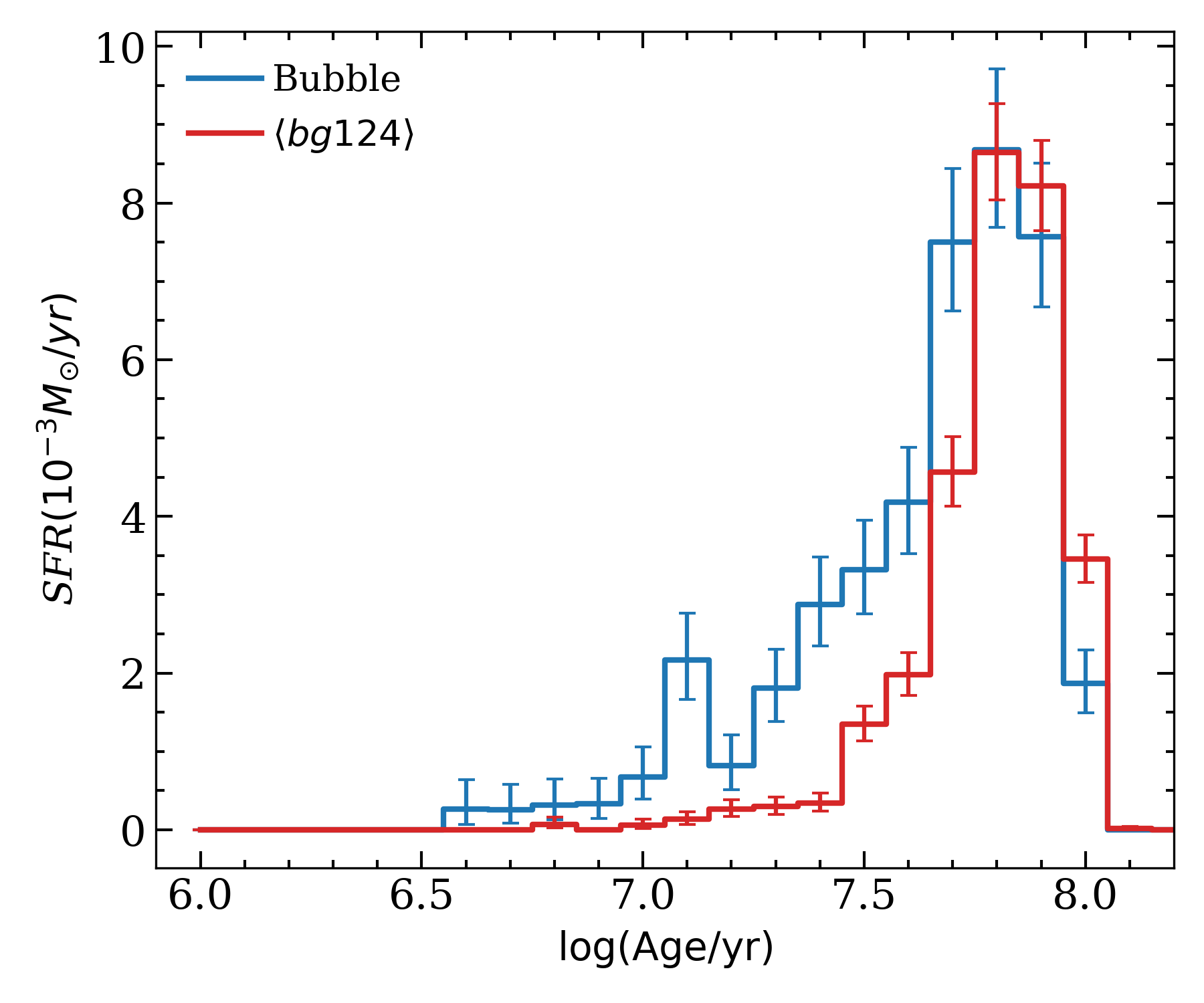}
    \caption{Normalized SFH for the bubble (blue) and disk (red) samples.
    There is a clear excess SFR inside the bubble between 10 and 50~Myr. 
    }
    \label{fig:sfh.parsec}
\end{figure}

\begin{figure}
    \centering
    \includegraphics[width=0.45\textwidth]{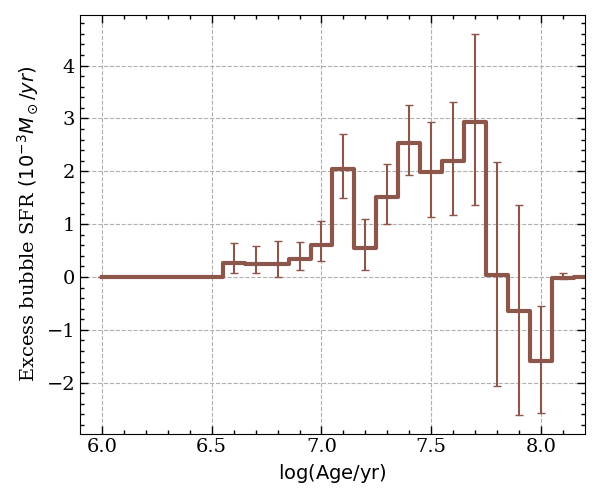}
    \includegraphics[width=0.45\textwidth]{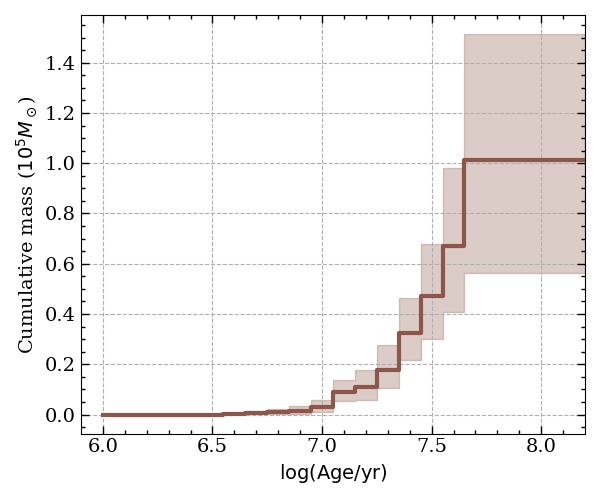}
    \caption{(Top) Excess SFR in the bubble with respect to the surrounding disk SFR.
    (Bottom) Cumulative mass of the stars formed inside the bubble.
    }
    \label{fig:res.sfh.parsec}
\end{figure}

\subsubsection{Age of the bubble population}

We analyse with our Bayesian code the star samples of the bubble and the four disk regions described in Sec.~\ref{sec:Sextractor}. Instead of cleaning disk stars in the bubble CMD, we infer the SFH of each sample independently, and then compare the relative differences in SFH between the bubble and the backgrounds.
We found that the differences in the numbers of stars in the bg1, bg2 and bg4 samples are less than the expected Poissonian error, whereas the bg3 sample contains distinctly less number of stars. In order to increase the statistical significance of the results, we determine the SFH of the disk for a sample obtained by combining the bg1, bg2 and bg4 disk samples, which we refer to as bg124 sample henceforth, and dividing the resulting SFH by a factor of three, which we denote by $<{\rm bg124}>$. We correct the bubble and the $<$bg124$>$ samples for the foreground Galactic extinction, using $\av=0.19$ mag \citep{Schlafly_2011}.

The SFHs of the bubble and the $<$bg124$>$ samples are shown in Fig.\,\ref{fig:sfh.parsec}. There is a clear excess SF in the bubble over the 50~Myr. In order to demonstrate this excess star formation, we subtract the normalized SFR obtained for the combined disk sample from that for the bubble sample, which is shown in the top panel of Figure~\ref{fig:res.sfh.parsec}. 
These plots illustrate that the SF started inside the bubble around 50~Myr ago, with the SFR decreasing steadily thereafter. The decrease in SF is more drastic over the last 10~Myr with no new stars formed in the most recent 4~Myr. The surrounding regions in the disk are also forming stars in the last 4--50~Myr, but at a rate smaller than that inside the bubble. 

\subsubsection{Mass of the bubble population}

In the bottom panel of Figure~\ref{fig:res.sfh.parsec} we show the cumulative mass of all the populations that contain an excess SFR in the bubble with respect to that in  the surrounding disk, which is obtained by integrating the excess SFR in the top panel.
The total mass of all stars formed inside the bubble over the last 50~Myr is $10^5$~\msol, of which 35\% was formed in the first 10~Myr (between 40 and 50~Myr ago). In comparison, only $\sim$20\% of stellar mass is formed in the most recent 20~Myr years. The remaining 45\% of stars were formed 20--40~Myr ago.

\section{Discussion}

\subsection{The energetics of the superbubble}

\begin{figure*}
\begin{centering}
\includegraphics[width=0.75\linewidth]{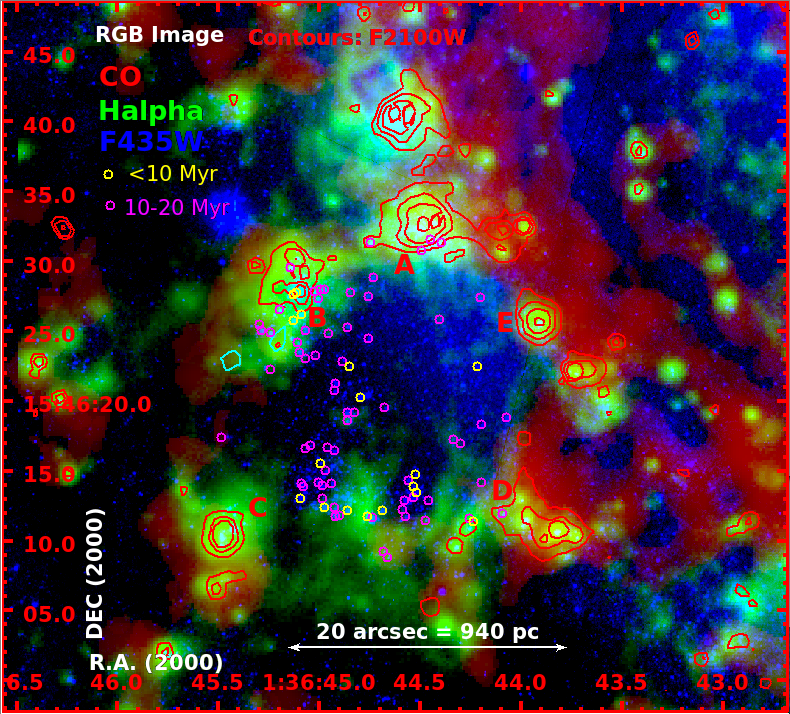}
\par\end{centering}
\caption{
An RGB image formed using CO (red), \ha\ (green) and F435W (blue) images. 
The red contours trace the bright knots seen in the F2100W filter band. The HST-detected stars younger than 20~Myr are identified with circles of yellow ($<$10~Myr) and magenta (10--20~Myr) colours. Unidentified stars (blue points) are older than 20~Myr.
The image illustrates that the star formation in the last 20~Myr has happened closer to the shell rather than the bubble center. The on-going star formation (regions marked by letters A to E), as traced by the F2100W contours, is happening at the densest parts of the expanding shell. See text for details. 
}
\label{fig:bubble_RGB}
\end{figure*}

\begin{figure}
\begin{centering}
\includegraphics[width=0.98\linewidth]{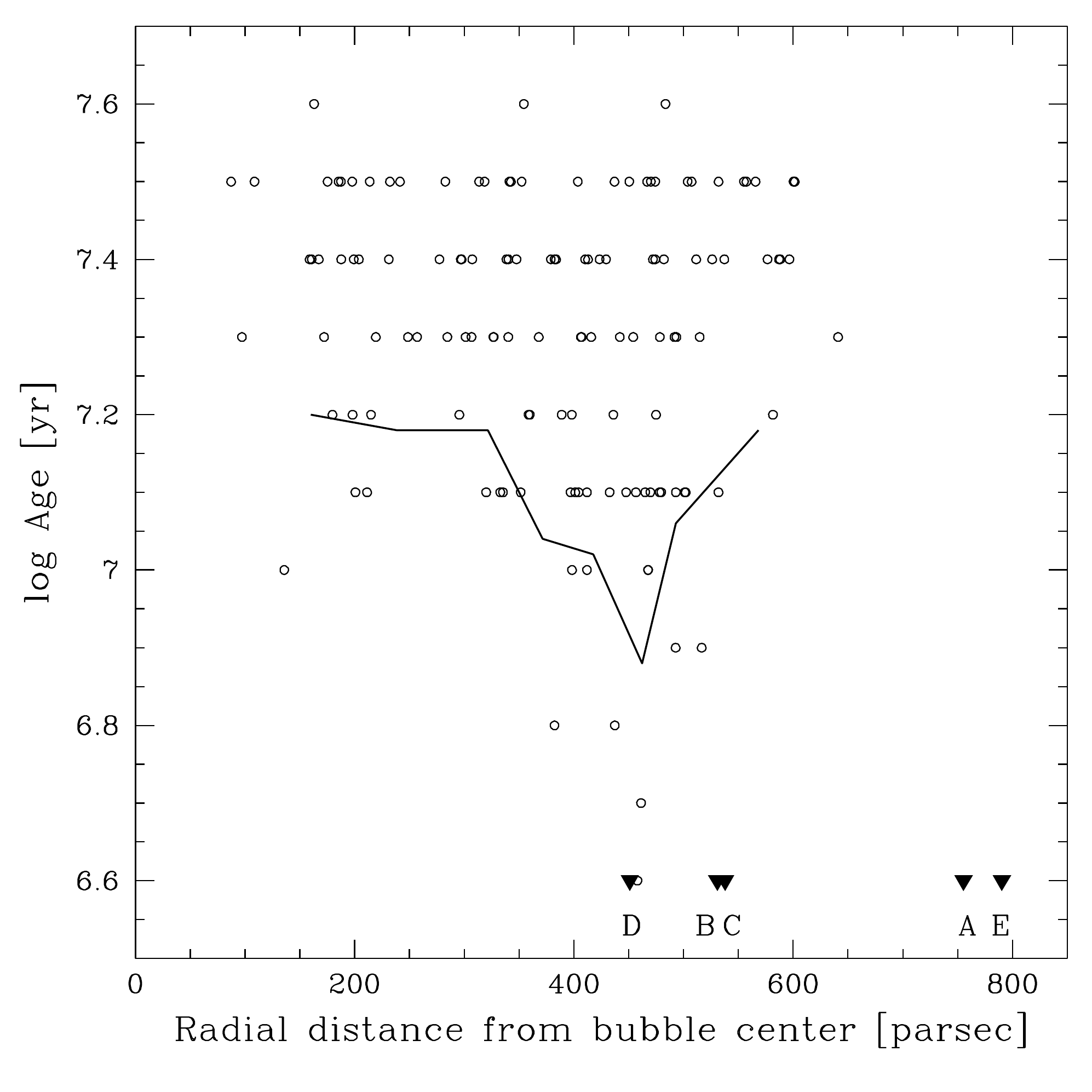}
\par\end{centering}
\caption{
Age of the stars inside the bubble plotted against their distance from the bubble center (circles). The solid line joins the age of the 5$^{\rm th}$ youngest star in successive radial bins. An upper age-limit of 4~Myr is assigned to the 21$\mu$m sources in the shell that are indicated by letters A to E. See text for details.
}
\label{fig:bubble_age_gradient}
\end{figure}

The creation and subsequent evolution of the superbubble by the correlated supernovae (SNe) explosions in disk galaxies has been studied in the classical work of \citet[][\cred{see also \cite{Bisnovatyi1995} and references therein}]{MacLow1988}. 
In the case of a homogeneous ambient gas distribution and a constant star cluster mechanical power  $L_\asterix$, the wind-driven superbubble radius $R_{\rm sb}$ and velocity $v_{\rm exp}$ at a time $t$ after the onset of expansion are given by,
\begin{equation}
R_{\rm sb} = A\left( \frac{L_\asterix}{\rho_{\rm gas}}\right)^{1/5} t^{3/5},
\label{eqn:hd1}
\end{equation}
\begin{equation}
v_{\rm exp} = \frac{3}{5}\frac{R_{\rm sb}}{t},
\label{eqn:hd2}
\end{equation}
where 
$\rho_{\rm gas}$ is the ambient gas density. Coefficient $A$ in Eq.~\ref{eqn:hd1} is:
\begin{equation}
A = \left[\frac{375(\gamma - 1)}{28(9\gamma-4)\pi}\right]^{1/5},
\label{eqn:hd3}
\end{equation}
where $\gamma\cred{=\frac{5}{3}}$ is the adiabatic index.
Eqn.~\ref{eqn:hd1} can be rewritten to solve for $L_\asterix$ in terms of observable parameters $R_{\rm sb}$, $v_{\rm exp}$ and $M_{\rm sh}$, the mass of the star cluster wind-driven shell. To do that, we first eliminate $t$ in Eqn.~\ref{eqn:hd1} by using Eqn.~\ref{eqn:hd2}:
\begin{equation}
L_\asterix = \frac{125}{27}\frac{\rho_{\rm gas} R_{\rm sb}^2 v_{\rm exp}^3}{A^5}.
\label{eqn:hd4}
\end{equation}
In the case of a homogeneous ambient medium $\rho_{\rm gas} = 3 M_{\rm sh}/4\pi R_{\rm sb}^3$. 
One can obtain then the cluster mechanical power as a function of the observed shell mass, radius and velocity:

\begin{equation}
L_\asterix = \frac{7(9\gamma-4)}{27(\gamma - 1)}\frac{M_{\rm sh} v_{\rm exp}^3}{R_{\rm sb}}.
\label{eqn:hd5}
\end{equation}
The value of the star cluster mechanical luminosity predicted by Eq.~\ref{eqn:hd5} could be confronted
with that obtained from the analysis of the JWST images.

In the case when $R_{\rm sb}$=450~pc, $M_{\rm sh}=2\times10^7$~\msol\ and $v_{\rm exp}$=12~\kms, 
\cred{the bubble age calculated using Eq.~\ref{eqn:hd2} is 22~Myr, which is consistent with the age of the enclosed stellar populations.
The mechanical power required to create the bubble using Eq.~\ref{eqn:hd5} is $L_{\rm mech}=2.2\times 10^{38}$ \ergs. The mechanical power injected into the ISM by the winds of O and WR stars and SN explosions in a  stellar population of $1\times10^5$~\msol\ at  solar metallicity is $L=5.0\times 10^{39}$ \ergs\ \citep{SB99}, which is
$\sim$20 times more than the mechanical power required to create the observed bubble in a homogeneous ISM.
}
\cred{
We cannot exclude, however, the possibility that the bubble tops were accelerated and perforated via Rayleigh-Taylor instabilities, and that some fraction of the deposited energy has escaped into the galaxy halo as the 450~pc inner radius of the shell exceeds the typical scale height of the gas disk \citep[see][and references therein]{MacLow1989, Tenorio-Tagle1990, Silich1998, Lopez2011}. In this case, the swept-up shell expansion is maintained by the injected momentum rather than energy. This agrees with the fact that {\sc Chandra} and {\sc XMM-Newton} observations did not reveal a significant amount of diffuse X-ray emission from the discussed region \citep{Soria2002}. A more comprehensive bubble model that includes the disk-like ambient gas distribution and non-coeval star formation and a more realistic spatial distribution of stars within the bubble will be presented elsewhere.
}

\subsection{Star formation scenario inside the bubble}

Our analysis of the bubble stars in the CMD has indicated that the bubble contains stars that are formed over the last 50~Myr, which contribute to total stellar mass of $10^5$~\msol. 
We here analyze the spatial distribution of stars of different ages in order to understand the star formation scenario inside the bubble since its formation.

In Figure~\ref{fig:bubble_RGB}, we show the distribution of the brightest of the stars inside the bubble in an RGB image formed by combining CO, \ha\ and F435W, respectively as red, green and blue components. In addition, we show the compact sources detected in the F2100W images by the contours. There is one-to-one correspondence between the locations of these 21~$\mu$m\ sources and the compact \ha\ sources. We identify five of the most prominent of these common sources in F2100W and \ha\ images by letters A, B, C, D and E. The youngest stars are identified by circles of yellow ($<$10~Myr) and magenta (10--20~Myr) colours. We do not indicate stars older than 20~Myr in the image to avoid congestion of points. 

If stars comprising each population that was inferred from the analysis of SFH belonged to star clusters, we would have expected agglomeration of stars of same age (i.e. circles of same colour in the figure). No such age-dependent spatial clustering of stars is easily discernible. Nevertheless, a closer inspection reveals that the 10--20~Myr old stars (magenta circles) are not distributed uniformly throughout the bubble, and instead majority of them are concentrated in three recognizable groups.
Two of these groups are located at the southern edge of the cavity between the regions identified as C and D. Both these groups are \cred{surrounded} by their own tiny \ha\ bubbles of $\sim$100~pc radius. The third stellar subgroup is situated closer to the center of the bubble to the south of the region B. Additionally, there is a crowded distribution of stars older than 20~Myr towards the CO-bright arc to the north-west. 
The distribution suggests that the younger stars are preferentially located near the shell rather than at the bubble center. 

In order to demonstrate an age-dependent spatial segregation of stars, we plot in Figure~\ref{fig:bubble_age_gradient}, the radial distance from the bubble center of each star associated to a particular isochrone. The stars associated with older isochrones are found at all radial distances,  whereas stars younger than 12~Myr are preferentially found at larger radial distances. We binned the data into 12 radial bins, each containing 16 stars, and joined with a solid line the age corresponding to the 5$^{\rm th}$ youngest star in each bin. The age of the youngest population is systematically lower at larger radii up to the ring radius of 450~pc. This suggests that the zone of active star formation is systematically moving away from the center as the bubble expands. It is most likely that the new stars are formed from the material that was swept up by the expanding bubble. 

There is a clear evidence for such a triggered star formation further out in the shell, which marks the location of the swept-up \cred{gas}, in the form of the sources identified by letters A to E. As described above, these sources are traced in \ha\ and 21~$\mu$m, with the former emission suggesting the presence of an ionizing star at these locations, and the latter suggesting the presence of hot dust surrounding that star. Faint F435W sources are detected at the location of these knots. Unfortunately, only region B is captured within the area over which we analysed ages. We infer an age of $<$10~Myr for two of the HST stars coinciding with the peak of the F2100W emission. Thus, it seems that the F2100W sources are not heavily embedded, in which case no stars in the HST images, and the \ha\ emission would have been detected at these locations. 
The newly formed stars are expected to get rid of the obscuring dust, and become undetectable in the F2100W filter in a few million years. 
All these characteristics suggest that the  sources A to E trace the most recent star formation event and we assign them an upper age limit of 4~Myr. Correspondingly, we show these sources in the figure by downward-pointing triangles, each identified with their names, at fixed upper age limit of 4~Myr. These points continue the trend seen inside the bubble of star formation progressively shifting to larger radius with time.

A total of $\sim10^5$~\msol\ of stars have been formed inside the bubble over the last 50~Myr in several star formation episodes. Each episode is expected to be associated with one or more star clusters of size not exceeding $\sim$10~pc, the typical size of open clusters. The stellar groups discussed above are the only identifiable agglomerations, with most of the stars inside the bubble  distributed over almost a kiloparsec scale. 
The logical explanation for this is that the stars were formed in clusters but the cluster got disrupted quickly.
Clusters are prone to disruption when the gas in the cluster volume gets expelled due to the mechanical power input from the massive stars in the form of stellar winds and supernova explosions \citep[e.g][]{Krause2016}. This starts to happen at $\sim$4~Myr. 
Cluster is expected to expand and survive if the expelled gas mass forms only a small fraction of the total mass within the cluster volume. The fraction of gas mass in a cluster depends on the star formation efficiency (SFE), with less residual gas for more efficient SF. According to hydrodynamical calculations of \citet{Krause2016}, a $10^5$~\msol\ cluster produces enough mechanical power to expel all its gas if it is not very compact (e.g. $r_h$=10~pc), even for large residual gas masses (or low SFE). It is likely that the progenitor cluster that was present at the center of the present-day bubble was not very compact and the binding energy of the cluster was not strong enough to counter the pressure exerted by its stars, following the expulsion of residual gas. If the cluster was in Virial equilibrium before it got dissolved, cluster stars at the most would have had a dispersion velocity $v_{\rm disp}$=10~\kms.

The currently observed spatial extent of the dissolved cluster allows us to check the above hypothesis. After dissolution, cluster stars are expected to move radially outward with a mean velocity which should be equivalent to the dispersion velocity of the progenitor cluster. At the inferred $v_{\rm disp}$=10~\kms, stars would reach 500~pc in 50~Myr. Thus those formed in the first episode had enough time to occupy the entire area of the present bubble. Stars that are formed at subsequent  epochs formed at some location between the bubble center and the shell, in such a way that these stars also would have had time to reach the boundaries of the bubble. Thus, the large spatial extent of the recently-formed stars inside the bubble can be reconciled with the scenario of disruption driven by gas expulsion and subsequent movement of the stars with the velocities that they had  when they were bound to the cluster. 

While the scenario described above explains the reason for the distribution of stars over the kiloparsec scale, it does not satisfactorily explain the observed complete absence of stars outside the bubble, and the strict confinement of stars inside the bubble. We here propose an alternative scenario, known as the gravitational feedback, that consistently explains the cluster disruption, \cred{and} the confinement of stars inside the bubble. The gravitational feedback is the process in which the gravitational potential well of the shell gas is deeper than that of the cluster, which not only destabilizes the cluster, but also accelerates the stars towards the shell. \citet{Zamora2019} showed that this happens in expanding bubbles when the  swept-up mass is orders of magnitude more than the cluster mass. They found that the ``Hubble-flow'' nature of the observed velocity fields of stars in some of the Galactic open clusters \citep[see e.g.][]{Kounkel2017, Roman2019} can be explained by the process.
The  gravitational force exerted by the mass collected in the expanding ring not only pulls apart the stars in an initially bound cluster, but also accelerates the stars towards the shell. 

\subsection{The conditions for the formation of kiloparsec-size bubbles}

In the analysis carried out in this work, we comprehensively showed that the $\sim10^5$~\msol\ stars formed in multiple episodes over the last 50~Myr, and the feedback from these stars, are responsible for the observed characteristics of the bubble. Such star-forming events are  common in galaxies unlike the presence of a kiloparsec size bubble. So, what is the most crucial  ingredient for the creation of such large bubbles?  

\citet{Vorobyov2005} found that the bubbles created by stellar feedback from multiple star formation events sustain a kiloparsec-diameter  organized bubble as compared to the one created by a massive cluster or exotic events like a hypernova explosion associated with a gamma ray burst \citep[GRB;][]{Efremov1998}, and the vertical impact of a high velocity cloud \citep[HVC;][]{Tenorio-Tagle1987}, all of the same total energy output. Their computations suggest that the location and mass of the first-formed cluster in the NGC628 bubble was the most crucial ingredient to create the observed large bubble. The SFR in the last bin in Figure~\ref{fig:res.sfh.parsec} suggests a mass of at the most  $3.5\times10^4$~\msol\ of stellar mass in the first cluster that was formed. The location of this cluster in the interarm region provided conditions for a uniform expansion of the bubble in its initial stages, which then formed new generations of stars in the shell \cred{that was} formed from the swept-up material. Given the lower ISM density in the interarm region as compared to a spiral arm, the cluster formed in the swept-up \cred{gas} mass was also of relatively smaller mass, thus allowing for a uniform, rather than a violent, expansion of the bubble. This process of steady expansion, collect and collapse, continued in each subsequent star formation event, each time forming moderate mass clusters. The bubble at the current time has reached the location of the spiral arm in the north-west side, where the bubble expansion seems to have been stalled. In the diagonally opposite direction, the bubble is growing asymmetrically and is less defined. We believe the bubble has reached the maximum size over which it can maintain a coherent structure.

\section{Conclusions}

We analysed the multiband data for a kiloparsec size bubble in the late type spiral galaxy NGC\,628, to understand the creation and maintainance of large ISM bubbles. The bubble is the largest of many such structures seen in the JWST F770W  images of this and other late-type galaxies, which  has given them the popular name of ``phantom galaxies''. The bubble seen in the F770W filter corresponds to the PAH molecules in the diffuse ISM, and is detected in other tracers of multi-phase ISM, such as in molecular gas by the CO emission, neutral atomic gas by the \hi\ line and ionized gas by the \ha\ emission. A kinematical analysis of the CO velocity field reveals that the bubble is currently expanding at a velocity of $\sim$12\,\kms. The mass of the gas in the shell around the bubble is $2\times10^7$~\msol, which is mostly in molecular form.
We find a clear evidence for the presence of a spatially resolved stellar population that is confined to the inner boundaries of bubble in the HST images. An isochrone-fitting analysis of these stars in the colour-magnitude diagram using a Bayesian technique, reveals that they are formed in multiple star formation events over the last 50~Myr. We find evidence for age-dependent spatial segregation of star formation with recent star formation happening closer to the inner shell of the bubble, rather than in its center, and on-going star formation at the shell itself. We propose a scenario wherein star formation started  at the center of the present bubble with the formation of a star cluster of $10^4$~\msol\, which got dissolved when the gas in the cluster volume got expelled  due to the combined energy output of stellar winds and supernovae as the massive stars in the cluster ended their lives. The mechanical power created by these massive stars gave rise to an expanding bubble. The swept-up mass in the shell enclosing the expanding bubble got dense enough to form new generations of stars. This cycle of cluster formation, its destruction, subsequent creation of an expanding shell, and secondary star formation in the shell gas continued until the present day, in the process forming a large bubble. We propose that the location of the seed cluster in the interarm region, and its relatively low mass are the critical ingredients required to set in motion the creation of a kiloparsec size interstellar bubble. 

\section*{Acknowledgements}
We thank the PHANGS team for making  the ALMA processed data publicly available. This paper makes use of the following ALMA data: ADS/JAO.ALMA\#2012.1.00650.S.
\cred{We thank the referee for the comments that have contributed to improvements of the manuscript.}
We also thank CONACyT for the research grant CB-A1-S-25070(YDM), A1-S-28458 (SS), CB-A1-S-22784 (DRG) and CF2022-320152 (DFA).
JAA acknowledges support from the IRyA-UNAM computer department and for granting him the required CPU time in the computers acquired through CONACyT grant CB2015-252364.
LLN thanks Funda\c{c}\~ao de Amparo\`a Pesquisa do Estado do Rio de Janeiro (FAPERJ) for granting the postdoctoral research fellowship E-40/2021(280692). 

\section{Data availability}

The paper is based on publicly available archival data.
 The processed fits files and data tables will be shared on reasonable request to the first author.

\bibliographystyle{mnras}
\bibliography{references} 

\begin{thebibliography}{}
\makeatletter
\relax
\def\mn@urlcharsother{\let\do\@makeother \do\$\do\&\do\#\do\^\do\_\do\%\do\~}
\def\mn@doi{\begingroup\mn@urlcharsother \@ifnextchar [ {\mn@doi@}
  {\mn@doi@[]}}
\def\mn@doi@[#1]#2{\def\@tempa{#1}\ifx\@tempa\@empty \href
  {http://dx.doi.org/#2} {doi:#2}\else \href {http://dx.doi.org/#2} {#1}\fi
  \endgroup}
\def\mn@eprint#1#2{\mn@eprint@#1:#2::\@nil}
\def\mn@eprint@arXiv#1{\href {http://arxiv.org/abs/#1} {{\tt arXiv:#1}}}
\def\mn@eprint@dblp#1{\href {http://dblp.uni-trier.de/rec/bibtex/#1.xml}
  {dblp:#1}}
\def\mn@eprint@#1:#2:#3:#4\@nil{\def\@tempa {#1}\def\@tempb {#2}\def\@tempc
  {#3}\ifx \@tempc \@empty \let \@tempc \@tempb \let \@tempb \@tempa \fi \ifx
  \@tempb \@empty \def\@tempb {arXiv}\fi \@ifundefined
  {mn@eprint@\@tempb}{\@tempb:\@tempc}{\expandafter \expandafter \csname
  mn@eprint@\@tempb\endcsname \expandafter{\@tempc}}}

\bibitem[\protect\citeauthoryear{{Alzate}, {Bruzual}  \&
  {D{\'\i}az-Gonz{\'a}lez}}{{Alzate} et~al.}{2021}]{Alzate2021}
{Alzate} J.~A.,  {Bruzual} G.,   {D{\'\i}az-Gonz{\'a}lez} D.~J.,  2021, \mn@doi
  [\mnras] {10.1093/mnras/staa3576}, \href
  {https://ui.adsabs.harvard.edu/abs/2021MNRAS.501..302A} {501, 302}

\bibitem[\protect\citeauthoryear{{Bertin} \& {Arnouts}}{{Bertin} \&
  {Arnouts}}{1996}]{Bertin1996}
{Bertin} E.,  {Arnouts} S.,  1996, \mn@doi [\aaps] {10.1051/aas:1996164}, \href
  {https://ui.adsabs.harvard.edu/abs/1996A&AS..117..393B} {117, 393}

\bibitem[\protect\citeauthoryear{{Bisnovatyi-Kogan} \&
  {Silich}}{{Bisnovatyi-Kogan} \& {Silich}}{1995}]{Bisnovatyi1995}
{Bisnovatyi-Kogan} G.~S.,  {Silich} S.~A.,  1995, \mn@doi [Reviews of Modern
  Physics] {10.1103/RevModPhys.67.661}, \href
  {https://ui.adsabs.harvard.edu/abs/1995RvMP...67..661B} {67, 661}

\bibitem[\protect\citeauthoryear{{Bolatto}, {Wolfire}  \& {Leroy}}{{Bolatto}
  et~al.}{2013}]{2013ARA&A..51..207B}
{Bolatto} A.~D.,  {Wolfire} M.,   {Leroy} A.~K.,  2013, \mn@doi [\araa]
  {10.1146/annurev-astro-082812-140944}, \href
  {https://ui.adsabs.harvard.edu/abs/2013ARA&A..51..207B} {51, 207}

\bibitem[\protect\citeauthoryear{{Boomsma}, {Oosterloo}, {Fraternali}, {van der
  Hulst}  \& {Sancisi}}{{Boomsma} et~al.}{2008}]{Boomsma2008}
{Boomsma} R.,  {Oosterloo} T.~A.,  {Fraternali} F.,  {van der Hulst} J.~M.,
  {Sancisi} R.,  2008, \mn@doi [\aap] {10.1051/0004-6361:200810120}, \href
  {https://ui.adsabs.harvard.edu/abs/2008A&A...490..555B} {490, 555}

\bibitem[\protect\citeauthoryear{{Bouchet} et~al.,}{{Bouchet}
  et~al.}{2015}]{Bouchet2015}
{Bouchet} P.,  et~al., 2015, \mn@doi [\pasp] {10.1086/682254}, \href
  {https://ui.adsabs.harvard.edu/abs/2015PASP..127..612B} {127, 612}

\bibitem[\protect\citeauthoryear{{Bressan}, {Marigo}, {Girardi}, {Salasnich},
  {Dal Cero}, {Rubele}  \& {Nanni}}{{Bressan} et~al.}{2012}]{Bressan2012}
{Bressan} A.,  {Marigo} P.,  {Girardi} L.,  {Salasnich} B.,  {Dal Cero} C.,
  {Rubele} S.,   {Nanni} A.,  2012, \mn@doi [\mnras]
  {10.1111/j.1365-2966.2012.21948.x}, \href
  {https://ui.adsabs.harvard.edu/abs/2012MNRAS.427..127B} {427, 127}

\bibitem[\protect\citeauthoryear{{Brinks} \& {Bajaja}}{{Brinks} \&
  {Bajaja}}{1986}]{Brinks1986}
{Brinks} E.,  {Bajaja} E.,  1986, \aap, \href
  {https://ui.adsabs.harvard.edu/abs/1986A&A...169...14B} {169, 14}

\bibitem[\protect\citeauthoryear{{Chen}, {Bressan}, {Girardi}, {Marigo}, {Kong}
   \& {Lanza}}{{Chen} et~al.}{2015}]{Chen2015}
{Chen} Y.,  {Bressan} A.,  {Girardi} L.,  {Marigo} P.,  {Kong} X.,   {Lanza}
  A.,  2015, \mn@doi [\mnras] {10.1093/mnras/stv1281}, \href
  {https://ui.adsabs.harvard.edu/abs/2015MNRAS.452.1068C} {452, 1068}

\bibitem[\protect\citeauthoryear{{Daigle}, {Carignan}, {Amram}, {Hernandez},
  {Chemin}, {Balkowski}  \& {Kennicutt}}{{Daigle}
  et~al.}{2006}]{2006MNRAS.367..469D}
{Daigle} O.,  {Carignan} C.,  {Amram} P.,  {Hernandez} O.,  {Chemin} L.,
  {Balkowski} C.,   {Kennicutt} R.,  2006, \mn@doi [\mnras]
  {10.1111/j.1365-2966.2006.10002.x}, \href
  {https://ui.adsabs.harvard.edu/abs/2006MNRAS.367..469D} {367, 469}

\bibitem[\protect\citeauthoryear{{De Young} \& {Heckman}}{{De Young} \&
  {Heckman}}{1994}]{DeYoung1994}
{De Young} D.~S.,  {Heckman} T.~M.,  1994, \mn@doi [\apj] {10.1086/174510},
  \href {https://ui.adsabs.harvard.edu/abs/1994ApJ...431..598D} {431, 598}

\bibitem[\protect\citeauthoryear{{Efremov}, {Elmegreen}  \& {Hodge}}{{Efremov}
  et~al.}{1998}]{Efremov1998}
{Efremov} Y.~N.,  {Elmegreen} B.~G.,   {Hodge} P.~W.,  1998, \mn@doi [\apjl]
  {10.1086/311468}, \href
  {https://ui.adsabs.harvard.edu/abs/1998ApJ...501L.163E} {501, L163}

\bibitem[\protect\citeauthoryear{{Emsellem} et~al.,}{{Emsellem}
  et~al.}{2022}]{Emsellem2022}
{Emsellem} E.,  et~al., 2022, \mn@doi [\aap] {10.1051/0004-6361/202141727},
  \href {https://ui.adsabs.harvard.edu/abs/2022A&A...659A.191E} {659, A191}

\bibitem[\protect\citeauthoryear{Gelman, Carlin, Stern, Dunson, Vehtari  \&
  Rubin}{Gelman et~al.}{2013}]{gelman13}
Gelman A.,  Carlin J.,  Stern H.,  Dunson D.,  Vehtari A.,   Rubin D.,  2013,
  Bayesian Data Analysis, Third Edition.
Chapman \& Hall/CRC Texts in Statistical Science, Taylor \& Francis, \url
  {https://books.google.com.mx/books?id=ZXL6AQAAQBAJ}

\bibitem[\protect\citeauthoryear{{Hensley}, {Murray}  \& {Dodici}}{{Hensley}
  et~al.}{2022}]{Hensley2022}
{Hensley} B.~S.,  {Murray} C.~E.,   {Dodici} M.,  2022, \mn@doi [\apj]
  {10.3847/1538-4357/ac5cbd}, \href
  {https://ui.adsabs.harvard.edu/abs/2022ApJ...929...23H} {929, 23}

\bibitem[\protect\citeauthoryear{{Hunter} \& {Gallagher}}{{Hunter} \&
  {Gallagher}}{1990}]{Hunter1990}
{Hunter} D.~A.,  {Gallagher} John~S. I.,  1990, \mn@doi [\apj]
  {10.1086/169286}, \href
  {https://ui.adsabs.harvard.edu/abs/1990ApJ...362..480H} {362, 480}

\bibitem[\protect\citeauthoryear{{Koo} \& {McKee}}{{Koo} \&
  {McKee}}{1992}]{1992ApJ...388...93K}
{Koo} B.-C.,  {McKee} C.~F.,  1992, \mn@doi [\apj] {10.1086/171132}, \href
  {https://ui.adsabs.harvard.edu/abs/1992ApJ...388...93K} {388, 93}

\bibitem[\protect\citeauthoryear{{Kounkel} et~al.,}{{Kounkel}
  et~al.}{2017}]{Kounkel2017}
{Kounkel} M.,  et~al., 2017, \mn@doi [\apj] {10.3847/1538-4357/834/2/142},
  \href {https://ui.adsabs.harvard.edu/abs/2017ApJ...834..142K} {834, 142}

\bibitem[\protect\citeauthoryear{{Krause}, {Charbonnel}, {Bastian}  \&
  {Diehl}}{{Krause} et~al.}{2016}]{Krause2016}
{Krause} M. G.~H.,  {Charbonnel} C.,  {Bastian} N.,   {Diehl} R.,  2016,
  \mn@doi [\aap] {10.1051/0004-6361/201526685}, \href
  {https://ui.adsabs.harvard.edu/abs/2016A&A...587A..53K} {587, A53}

\bibitem[\protect\citeauthoryear{{Kreckel} et~al.,}{{Kreckel}
  et~al.}{2018}]{2018ApJ...863L..21K}
{Kreckel} K.,  et~al., 2018, \mn@doi [\apjl] {10.3847/2041-8213/aad77d}, \href
  {https://ui.adsabs.harvard.edu/abs/2018ApJ...863L..21K} {863, L21}

\bibitem[\protect\citeauthoryear{{Kreckel} et~al.,}{{Kreckel}
  et~al.}{2019}]{Kreckel2019}
{Kreckel} K.,  et~al., 2019, \mn@doi [\apj] {10.3847/1538-4357/ab5115}, \href
  {https://ui.adsabs.harvard.edu/abs/2019ApJ...887...80K} {887, 80}

\bibitem[\protect\citeauthoryear{{Kroupa}}{{Kroupa}}{2001}]{Kroupa2001}
{Kroupa} P.,  2001, \mn@doi [\mnras] {10.1046/j.1365-8711.2001.04022.x}, \href
  {https://ui.adsabs.harvard.edu/abs/2001MNRAS.322..231K} {322, 231}

\bibitem[\protect\citeauthoryear{{Kuhn}, {Hillenbrand}, {Sills}, {Feigelson}
  \& {Getman}}{{Kuhn} et~al.}{2019}]{2019ApJ...870...32K}
{Kuhn} M.~A.,  {Hillenbrand} L.~A.,  {Sills} A.,  {Feigelson} E.~D.,   {Getman}
  K.~V.,  2019, \mn@doi [\apj] {10.3847/1538-4357/aaef8c}, \href
  {https://ui.adsabs.harvard.edu/abs/2019ApJ...870...32K} {870, 32}

\bibitem[\protect\citeauthoryear{{Lada} \& {Lada}}{{Lada} \&
  {Lada}}{2003}]{Lada2003}
{Lada} C.~J.,  {Lada} E.~A.,  2003, \mn@doi [\araa]
  {10.1146/annurev.astro.41.011802.094844}, \href
  {https://ui.adsabs.harvard.edu/abs/2003ARA&A..41...57L} {41, 57}

\bibitem[\protect\citeauthoryear{{Leitherer} et~al.,}{{Leitherer}
  et~al.}{1999}]{SB99}
{Leitherer} C.,  et~al., 1999, \mn@doi [\apjs] {10.1086/313233}, \href
  {https://ui.adsabs.harvard.edu/abs/1999ApJS..123....3L} {123, 3}

\bibitem[\protect\citeauthoryear{{Leroy} et~al.,}{{Leroy}
  et~al.}{2009}]{2009AJ....137.4670L}
{Leroy} A.~K.,  et~al., 2009, \mn@doi [\aj] {10.1088/0004-6256/137/6/4670},
  \href {https://ui.adsabs.harvard.edu/abs/2009AJ....137.4670L} {137, 4670}

\bibitem[\protect\citeauthoryear{{Leroy} et~al.,}{{Leroy}
  et~al.}{2021a}]{2021ApJS..255...19L}
{Leroy} A.~K.,  et~al., 2021a, \mn@doi [\apjs] {10.3847/1538-4365/abec80},
  \href {https://ui.adsabs.harvard.edu/abs/2021ApJS..255...19L} {255, 19}

\bibitem[\protect\citeauthoryear{{Leroy} et~al.,}{{Leroy}
  et~al.}{2021b}]{2021ApJS..257...43L}
{Leroy} A.~K.,  et~al., 2021b, \mn@doi [\apjs] {10.3847/1538-4365/ac17f3},
  \href {https://ui.adsabs.harvard.edu/abs/2021ApJS..257...43L} {257, 43}

\bibitem[\protect\citeauthoryear{{Lopez}, {Krumholz}, {Bolatto}, {Prochaska}
  \& {Ramirez-Ruiz}}{{Lopez} et~al.}{2011}]{Lopez2011}
{Lopez} L.~A.,  {Krumholz} M.~R.,  {Bolatto} A.~D.,  {Prochaska} J.~X.,
  {Ramirez-Ruiz} E.,  2011, \mn@doi [\apj] {10.1088/0004-637X/731/2/91}, \href
  {https://ui.adsabs.harvard.edu/abs/2011ApJ...731...91L} {731, 91}

\bibitem[\protect\citeauthoryear{{Lu}, {Hoffman}, {Groff}, {Roos}  \&
  {Lamphier}}{{Lu} et~al.}{1993}]{1993ApJS...88..383L}
{Lu} N.~Y.,  {Hoffman} G.~L.,  {Groff} T.,  {Roos} T.,   {Lamphier} C.,  1993,
  \mn@doi [\apjs] {10.1086/191826}, \href
  {https://ui.adsabs.harvard.edu/abs/1993ApJS...88..383L} {88, 383}

\bibitem[\protect\citeauthoryear{{Mac Low} \& {Ferrara}}{{Mac Low} \&
  {Ferrara}}{1999}]{MacLow1999}
{Mac Low} M.-M.,  {Ferrara} A.,  1999, \mn@doi [\apj] {10.1086/306832}, \href
  {https://ui.adsabs.harvard.edu/abs/1999ApJ...513..142M} {513, 142}

\bibitem[\protect\citeauthoryear{{Mac Low} \& {McCray}}{{Mac Low} \&
  {McCray}}{1988}]{MacLow1988}
{Mac Low} M.-M.,  {McCray} R.,  1988, \mn@doi [\apj] {10.1086/165936}, \href
  {https://ui.adsabs.harvard.edu/abs/1988ApJ...324..776M} {324, 776}

\bibitem[\protect\citeauthoryear{{Mac Low}, {McCray}  \& {Norman}}{{Mac Low}
  et~al.}{1989}]{MacLow1989}
{Mac Low} M.-M.,  {McCray} R.,   {Norman} M.~L.,  1989, \mn@doi [\apj]
  {10.1086/167094}, \href
  {https://ui.adsabs.harvard.edu/abs/1989ApJ...337..141M} {337, 141}

\bibitem[\protect\citeauthoryear{{Micelotta}, {Jones}  \&
  {Tielens}}{{Micelotta} et~al.}{2010a}]{Micelotta2010a}
{Micelotta} E.~R.,  {Jones} A.~P.,   {Tielens} A.~G.~G.~M.,  2010a, \mn@doi
  [\aap] {10.1051/0004-6361/200911682}, \href
  {https://ui.adsabs.harvard.edu/abs/2010A&A...510A..36M} {510, A36}

\bibitem[\protect\citeauthoryear{{Micelotta}, {Jones}  \&
  {Tielens}}{{Micelotta} et~al.}{2010b}]{Micelotta2010b}
{Micelotta} E.~R.,  {Jones} A.~P.,   {Tielens} A.~G.~G.~M.,  2010b, \mn@doi
  [\aap] {10.1051/0004-6361/200911683}, \href
  {https://ui.adsabs.harvard.edu/abs/2010A&A...510A..37M} {510, A37}

\bibitem[\protect\citeauthoryear{{Montillaud}, {Joblin}  \&
  {Toublanc}}{{Montillaud} et~al.}{2013}]{Montillaud2013}
{Montillaud} J.,  {Joblin} C.,   {Toublanc} D.,  2013, \mn@doi [\aap]
  {10.1051/0004-6361/201220757}, \href
  {https://ui.adsabs.harvard.edu/abs/2013A&A...552A..15M} {552, A15}

\bibitem[\protect\citeauthoryear{{Olivares} et~al.,}{{Olivares}
  et~al.}{2010}]{Olivares2010}
{Olivares} F.,  et~al., 2010, \mn@doi [\apj] {10.1088/0004-637X/715/2/833},
  \href {https://ui.adsabs.harvard.edu/abs/2010ApJ...715..833O} {715, 833}

\bibitem[\protect\citeauthoryear{{Pabst} et~al.,}{{Pabst}
  et~al.}{2019}]{Pabst2019}
{Pabst} C.,  et~al., 2019, \mn@doi [\nat] {10.1038/s41586-018-0844-1}, \href
  {https://ui.adsabs.harvard.edu/abs/2019Natur.565..618P} {565, 618}

\bibitem[\protect\citeauthoryear{{Pabst} et~al.,}{{Pabst}
  et~al.}{2021}]{Pabst2021}
{Pabst} C.~H.~M.,  et~al., 2021, \mn@doi [\aap] {10.1051/0004-6361/202140804},
  \href {https://ui.adsabs.harvard.edu/abs/2021A&A...651A.111P} {651, A111}

\bibitem[\protect\citeauthoryear{{Pokhrel}, {Simpson}  \&
  {Bagetakos}}{{Pokhrel} et~al.}{2020}]{Pokhrel2020}
{Pokhrel} N.~R.,  {Simpson} C.~E.,   {Bagetakos} I.,  2020, \mn@doi [\aj]
  {10.3847/1538-3881/ab9bfa}, \href
  {https://ui.adsabs.harvard.edu/abs/2020AJ....160...66P} {160, 66}

\bibitem[\protect\citeauthoryear{{Rom{\'a}n-Z{\'u}{\~n}iga}, {Roman-Lopes},
  {Tapia}, {Hern{\'a}ndez}  \&
  {Ram{\'\i}rez-Preciado}}{{Rom{\'a}n-Z{\'u}{\~n}iga} et~al.}{2019}]{Roman2019}
{Rom{\'a}n-Z{\'u}{\~n}iga} C.~G.,  {Roman-Lopes} A.,  {Tapia} M.,
  {Hern{\'a}ndez} J.,   {Ram{\'\i}rez-Preciado} V.,  2019, \mn@doi [\apjl]
  {10.3847/2041-8213/aafb06}, \href
  {https://ui.adsabs.harvard.edu/abs/2019ApJ...871L..12R} {871, L12}

\bibitem[\protect\citeauthoryear{{Santoro} et~al.,}{{Santoro}
  et~al.}{2022}]{Santoro2022}
{Santoro} F.,  et~al., 2022, \mn@doi [\aap] {10.1051/0004-6361/202141907},
  \href {https://ui.adsabs.harvard.edu/abs/2022A&A...658A.188S} {658, A188}

\bibitem[\protect\citeauthoryear{Schlafly \& Finkbeiner}{Schlafly \&
  Finkbeiner}{2011}]{Schlafly_2011}
Schlafly E.~F.,  Finkbeiner D.~P.,  2011, \mn@doi [The Astrophysical Journal]
  {10.1088/0004-637X/737/2/103}, 737, 103

\bibitem[\protect\citeauthoryear{{Silich} \& {Tenorio-Tagle}}{{Silich} \&
  {Tenorio-Tagle}}{1998}]{Silich1998}
{Silich} S.~A.,  {Tenorio-Tagle} G.,  1998, \mn@doi [\mnras]
  {10.1046/j.1365-8711.1998.01765.x}, \href
  {https://ui.adsabs.harvard.edu/abs/1998MNRAS.299..249S} {299, 249}

\bibitem[\protect\citeauthoryear{{Silich}, {Tenorio-Tagle}, {Terlevich},
  {Terlevich}  \& {Netzer}}{{Silich} et~al.}{2001}]{Silich2001}
{Silich} S.~A.,  {Tenorio-Tagle} G.,  {Terlevich} R.,  {Terlevich} E.,
  {Netzer} H.,  2001, \mn@doi [\mnras] {10.1046/j.1365-8711.2001.04288.x},
  \href {https://ui.adsabs.harvard.edu/abs/2001MNRAS.324..191S} {324, 191}

\bibitem[\protect\citeauthoryear{{Small}, {Bersier}  \& {Salaris}}{{Small}
  et~al.}{2013}]{Small2013}
{Small} E.~E.,  {Bersier} D.,   {Salaris} M.,  2013, \mn@doi [\mnras]
  {10.1093/mnras/sts077}, \href
  {https://ui.adsabs.harvard.edu/abs/2013MNRAS.428..763S} {428, 763}

\bibitem[\protect\citeauthoryear{{Soria} \& {Kong}}{{Soria} \&
  {Kong}}{2002}]{Soria2002}
{Soria} R.,  {Kong} A. K.~H.,  2002, \mn@doi [\apjl] {10.1086/341445}, \href
  {https://ui.adsabs.harvard.edu/abs/2002ApJ...572L..33S} {572, L33}

\bibitem[\protect\citeauthoryear{{Steinwandel}, {Moster}, {Naab}, {Hu}  \&
  {Walch}}{{Steinwandel} et~al.}{2020}]{2020MNRAS.495.1035S}
{Steinwandel} U.~P.,  {Moster} B.~P.,  {Naab} T.,  {Hu} C.-Y.,   {Walch} S.,
  2020, \mn@doi [\mnras] {10.1093/mnras/staa821}, \href
  {https://ui.adsabs.harvard.edu/abs/2020MNRAS.495.1035S} {495, 1035}

\bibitem[\protect\citeauthoryear{{Tenorio-Tagle} \&
  {Bodenheimer}}{{Tenorio-Tagle} \& {Bodenheimer}}{1988}]{Tenorio-Tagle1988}
{Tenorio-Tagle} G.,  {Bodenheimer} P.,  1988, \mn@doi [\araa]
  {10.1146/annurev.aa.26.090188.001045}, \href
  {https://ui.adsabs.harvard.edu/abs/1988ARA&A..26..145T} {26, 145}

\bibitem[\protect\citeauthoryear{{Tenorio-Tagle}, {Franco}, {Bodenheimer}  \&
  {Rozyczka}}{{Tenorio-Tagle} et~al.}{1987}]{Tenorio-Tagle1987}
{Tenorio-Tagle} G.,  {Franco} J.,  {Bodenheimer} P.,   {Rozyczka} M.,  1987,
  \aap, \href {https://ui.adsabs.harvard.edu/abs/1987A&A...179..219T} {179,
  219}

\bibitem[\protect\citeauthoryear{{Tenorio-Tagle}, {Rozyczka}  \&
  {Bodenheimer}}{{Tenorio-Tagle} et~al.}{1990}]{Tenorio-Tagle1990}
{Tenorio-Tagle} G.,  {Rozyczka} M.,   {Bodenheimer} P.,  1990, \aap, \href
  {https://ui.adsabs.harvard.edu/abs/1990A&A...237..207T} {237, 207}

\bibitem[\protect\citeauthoryear{{Tielens}}{{Tielens}}{2008}]{Tielens2008}
{Tielens} A.~G.~G.~M.,  2008, \mn@doi [\araa]
  {10.1146/annurev.astro.46.060407.145211}, \href
  {https://ui.adsabs.harvard.edu/abs/2008ARA&A..46..289T} {46, 289}

\bibitem[\protect\citeauthoryear{{Vorobyov} \& {Basu}}{{Vorobyov} \&
  {Basu}}{2005}]{Vorobyov2005}
{Vorobyov} E.~I.,  {Basu} S.,  2005, \mn@doi [\aap]
  {10.1051/0004-6361:20041324}, \href
  {https://ui.adsabs.harvard.edu/abs/2005A&A...431..451V} {431, 451}

\bibitem[\protect\citeauthoryear{{Walmswell}, {Eldridge}, {Brewer}  \&
  {Tout}}{{Walmswell} et~al.}{2013}]{Walms13}
{Walmswell} J.~J.,  {Eldridge} J.~J.,  {Brewer} B.~J.,   {Tout} C.~A.,  2013,
  \mn@doi [\mnras] {10.1093/mnras/stt1444}, \href
  {https://ui.adsabs.harvard.edu/abs/2013MNRAS.435.2171W} {435, 2171}

\bibitem[\protect\citeauthoryear{{Walter}, {Brinks}, {de Blok}, {Bigiel},
  {Kennicutt}, {Thornley}  \& {Leroy}}{{Walter}
  et~al.}{2008}]{2008AJ....136.2563W}
{Walter} F.,  {Brinks} E.,  {de Blok} W.~J.~G.,  {Bigiel} F.,  {Kennicutt}
  Robert~C. J.,  {Thornley} M.~D.,   {Leroy} A.,  2008, \mn@doi [\aj]
  {10.1088/0004-6256/136/6/2563}, \href
  {https://ui.adsabs.harvard.edu/abs/2008AJ....136.2563W} {136, 2563}

\bibitem[\protect\citeauthoryear{{Warren} et~al.,}{{Warren}
  et~al.}{2011}]{Warren2011}
{Warren} S.~R.,  et~al., 2011, \mn@doi [\apj] {10.1088/0004-637X/738/1/10},
  \href {https://ui.adsabs.harvard.edu/abs/2011ApJ...738...10W} {738, 10}

\bibitem[\protect\citeauthoryear{{Wells} et~al.,}{{Wells}
  et~al.}{2015}]{Wells2015}
{Wells} M.,  et~al., 2015, \mn@doi [\pasp] {10.1086/682281}, \href
  {https://ui.adsabs.harvard.edu/abs/2015PASP..127..646W} {127, 646}

\bibitem[\protect\citeauthoryear{{Williams} et~al.,}{{Williams}
  et~al.}{2022}]{Williams2022}
{Williams} T.~G.,  et~al., 2022, \mn@doi [\apjl] {10.3847/2041-8213/aca674},
  \href {https://ui.adsabs.harvard.edu/abs/2022ApJ...941L..27W} {941, L27}

\bibitem[\protect\citeauthoryear{{Wright} et~al.,}{{Wright}
  et~al.}{2015}]{Wright2015}
{Wright} G.~S.,  et~al., 2015, \mn@doi [\pasp] {10.1086/682253}, \href
  {https://ui.adsabs.harvard.edu/abs/2015PASP..127..595W} {127, 595}

\bibitem[\protect\citeauthoryear{{Zamora-Avil{\'e}s}, {Ballesteros-Paredes},
  {Hern{\'a}ndez}, {Rom{\'a}n-Z{\'u}{\~n}iga}, {Lora}  \&
  {Kounkel}}{{Zamora-Avil{\'e}s} et~al.}{2019}]{Zamora2019}
{Zamora-Avil{\'e}s} M.,  {Ballesteros-Paredes} J.,  {Hern{\'a}ndez} J.,
  {Rom{\'a}n-Z{\'u}{\~n}iga} C.,  {Lora} V.,   {Kounkel} M.,  2019, \mn@doi
  [\mnras] {10.1093/mnras/stz1897}, \href
  {https://ui.adsabs.harvard.edu/abs/2019MNRAS.488.3406Z} {488, 3406}

\bibitem[\protect\citeauthoryear{{Zhang}, {Ho}  \& {Li}}{{Zhang}
  et~al.}{2022}]{Zhang2022}
{Zhang} L.,  {Ho} L.~C.,   {Li} A.,  2022, \mn@doi [\apj]
  {10.3847/1538-4357/ac930f}, \href
  {https://ui.adsabs.harvard.edu/abs/2022ApJ...939...22Z} {939, 22}

\bibitem[\protect\citeauthoryear{{van der Kruit} \& {Allen}}{{van der Kruit} \&
  {Allen}}{1978}]{1978ARA&A..16..103V}
{van der Kruit} P.~C.,  {Allen} R.~J.,  1978, \mn@doi [\araa]
  {10.1146/annurev.aa.16.090178.000535}, \href
  {https://ui.adsabs.harvard.edu/abs/1978ARA&A..16..103V} {16, 103}

\makeatother
\end{thebibliography}

\newpage

\appendix

\section{Definition of parameters used in the Bayesian method of age analysis}

\begin{table*} 
\begin{center}
 \caption{\label{tab:model_param}Variables entering the hierarchical model. Here dm, $A_{V}$ and $M$ are the distance modulus, the visual extinction and the stellar mass, respectively. 
 }
 \begin{tabular}{llcl}
 \hline
 Hyper-parameter                   &             $\bm{a}=\{a_{1},..,a_{N_{I}}\}$              &        &    Stellar fraction vector           \\
 \hline
  \multirow{2}{*}{Parameters}
                         &               $\mj$              &   mag  &    predicted absolute magnitude       \\
                         &               $\widehat{F}_{j}^{k}=\mj+{\rm dm}+A_{V}$              &   mag  &    predicted apparent magnitude       \\
 \hline
 \multirow{2}{*}{Data ($\bm{F}$)}
                                   &            $F_{j}^{k}$            &   mag  &    Observed apparent magnitude       \\
                                   &            $e_{j}^{k}$            &   mag  &    Apparent magnitude error          \\
 \hline
 \multirow{3}{*}{Fixed distributions}
                                   &        $\mathcal{F}_{i}^{k}(M)$      &   mag  &    Isochrone absolute magnitude      \\
                                   &          $\sigma_{i}^{k}$         &   mag  &    Isochrone tolerance       \\
                                   &              $\phi(M)$            &        &    Initial mass function             \\
 \hline
\end{tabular}
\end{center}
\end{table*}

\begin{table*}
\begin{center}
 \caption{\label{tab:model_pdf}Probability distribution function. Here $\Gamma$, $\xi$ are the Gamma function and the concentration parameter (See text).
 }
 \begin{tabular}{lcc}
 \hline
   Name            &              PDF              &                Eq.              \\
 \hline
 Hyperprior           &          \hpriorp          &            \hpriorpdf           \\
 \hline
 Likelihood        &            \likep             &            \likepdf             \\
 \hline
 Prior                &          \priorp           &            \priorpdf            \\
 \hline

\end{tabular}
\end{center}
\end{table*}

In Table~\ref{tab:model_param}, we give the mathematical definitions of random variables entering the hierarchical model.
Subscripts $i=1,2, ... , N_{I}$ and $j=1,2, ... , N_{D}$ indicate that the computation is made using the $i^{\rm th}$ isochrone and the $j^{\rm th}$ star,
where $N_{I}$ and $N_D$ are the number of isochrones and observed stars, respectively.
Superscript $k=1,2,3$ specifies the HST photometric magnitudes: $(f^{1},f^{2},f^{3})$=F435W, F555W, F814W.
The model establishes that the pdf of the predicted absolute magnitude $f^{k}$ is a function of the isochrone magnitudes $\mathcal{F}^{k}(M)$ as follows
\begin{equation}\label{eq:f.pdf}
f_{j}^{k}\sim \mathcal{N}(\mathcal{F}_{i}^{k},\sigma_{i}^{k}),
\end{equation}
taking the isochrone tolerance as constant, independent of $i$ and $k$.
Distance modulus and visual extinction, dm and $A_{V}$ respectively, are known a priory and they are not variables.

Table~\ref{tab:model_pdf}, we give the equations for the PDFs.
The hyper-prior $p(\bm{a})$ is the symmetric Dirichlet distribution, with concentration parameter $\xi$.
When $\xi=1$, the hyper-prior is uniform, and no $a_i$ is preferred.
See Sec. 3.4 of \citet{gelman13} for more details of the Dirichlet distribution, and Sec. 3.2 of \citet{Walms13} for an example of its application within an astronomical context.
Prior PDF is built as a mixture of model (isochrones) and its assemble the mass PDF and the $f^{k}$ PDF.

\end{document}